\documentclass[12pt]{article}

\usepackage{amssymb,amsmath,amsfonts,eurosym,geometry,ulem,graphicx,caption,color,setspace,sectsty,comment,footmisc,caption,natbib,pdflscape,subfigure,array}
\usepackage{tikz}
\usepackage{pgfplots}

\pgfplotsset{width=5cm,compat=1.9}

\usepgfplotslibrary{fillbetween}
\usepackage{color} 
\usepackage{hyperref}
\hypersetup{hidelinks,colorlinks=true,linkcolor=blue,citecolor=blue}

\usepackage{natbib}

\newcolumntype{L}[1]{>{\raggedright\let\newline\\arraybackslash\hspace{0pt}}m{#1}}
\newcolumntype{C}[1]{>{\centering\let\newline\\arraybackslash\hspace{0pt}}m{#1}}
\newcolumntype{R}[1]{>{\raggedleft\let\newline\\arraybackslash\hspace{0pt}}m{#1}}

\usepackage[T1]{fontenc}
\usepackage[utf8]{inputenc}
\usepackage{setspace}
\usepackage{babel}
\usepackage{blindtext}
\usepackage[utf8]{inputenc}
\usepackage{amssymb}
\usepackage{amsthm}
\usepackage{amsmath}
\newtheorem{theorem}{Theorem}
\newtheorem{corollary}{Corollary}
\newtheorem{example}{Example}
\newtheorem{proposition}{Proposition}

\newtheorem{remark}{Remark}
\newtheorem{lemma}{Lemma}

\usepackage{array}
\usepackage{footmisc}
\DefineFNsymbols{mySymbols}{{\ensuremath\dagger}{*}}
\setfnsymbol{mySymbols}

\newtheorem{definition}{Definition}

\title{Correlation-Robust Optimal Auctions}
\author{Wanchang Zhang\thanks{Department of Economics, University of California, San Diego. Email: waz024@ucsd.edu}\thanks{I am indebted to  Songzi Du and Joel Sobel for stimulating discussions. I thank Snehal Banerjee,  Bradyn Breon-Drish, Yi-Chun Chen,  Simone Galperti, Joel Watson and Xiangqian Yang  for helpful comments. }
}
\date{Current Draft: March 31,  2022\\First Draft: April 8, 2020}

\begin{document}

\maketitle
\begin{abstract}
   I study the design of auctions in which  the auctioneer is assumed to have  information only about the  marginal distribution of a generic bidder's valuation, but does not know
the correlation structure of the joint distribution of bidders' valuations. I assume that a generic bidder's valuation is bounded and $\bar{v}$ is the maximum valuation of a generic bidder.  The
performance of a mechanism is evaluated in the worst case over
the uncertainty of  joint distributions that are consistent with the marginal distribution. For the two-bidder case,  \textit{the second-price auction with the uniformly distributed random reserve} maximizes the worst-case expected revenue across \textit{all} dominant-strategy  mechanisms under certain regularity conditions. For the $N$-bidder ($N\ge 3$) case, \textit{the second-price auction with the $\bar{v}-$scaled $Beta (\frac{1}{N-1},1)$ distributed random reserve} maximizes the worst-case expected revenue across \textit{standard} (a bidder whose bid is not the highest will never be allocated) dominant-strategy  mechanisms under certain regularity conditions. When the probability mass condition  (part of the regularity conditions) does not hold,   \textit{the second-price auction with the $s^*-$scaled $Beta (\frac{1}{N-1},1)$ distributed random reserve} maximizes the worst-case expected revenue across standard dominant-strategy  mechanisms, where $s^*\in (0,\bar{v})$.  
\vspace{0in}\\
\noindent\textbf{Keywords:} Robust mechanism design, dominant-strategy mechanism, second-price auction, random reserve,  duality.\\
\vspace{0in}\\
\noindent\textbf{JEL Codes:} C72, D82, D83.\\

\bigskip

\end{abstract}
\newpage

\section{Introduction}\label{s1}
The  mechanism design literature assumes that  bidders'  valuation profile follows a \textit{commonly known} joint distribution. For example,   \cite{myerson1981optimal} assumes that the auctioneer knows the marginal distribution of each bidder's valuation, and also knows that bidders' valuations are independently distributed.  While the independent private value model is widely acknowledged as a useful benchmark, little is known about how the optimal mechanism would perform once the model is misspecified. In addition, it is not clear how the auctioneer should determine which model is the correct one to use.\\
\indent In this paper, I study the single-object auction problem in the correlated private value environment in which  bidders'  valuation
profile is drawn from a general joint
distribution. I assume
that the auctioneer knows the marginal distribution of a generic 
bidder's valuation, but does not have any
knowledge about the  correlation structure of different
bidders' valuations\footnote{The framework is originally
proposed by \cite{carroll2017robustness} for the multi-dimensional screening problem. His solution is simple and conveys a clear and intuitive message: if you do not know how to bundle, then do not. It is natural to adapt his framework to an environment with  multiple bidders whose private valuations may be correlated.}. A joint distribution of bidders' valuation profile is said to be \textit{conceivable} if it is consistent with the known marginal distribution of a generic bidder's valuation. The auctioneer seeks a dominant-strategy mechanism.  A mechanism is evaluated according
to the auctioneer’s expected revenue in the dominant-strategy equilibrium  derived in the worst case, referred to as the \textit{revenue guarantee},  over all conceivable joint distributions. The objective of the auctioneer  is to design a mechanism that maximizes the revenue guarantee across some general class of dominant-strategy mechanisms. I call such a mechanism a \textit{maxmin mechanism}. \\
\indent This framework is in the same spirit as the  robust mechanism literature in that it assumes away detailed knowledge of the auctioneer (\cite{wilson1987game}).    It is motivated by the observation that the joint distribution is a much higher-dimensional object than the marginal distribution of a generic bidder. Therefore it is more difficult to estimate the joint distribution.  Practically, it fits into the situations where the bidder pool changes constantly and then there is no data for estimating the correlation structure.  Another situation where the auctioneer may only
know the marginal distribution for each bidder is the one
where the identities of the participating bidders cannot be
observed. \\
\indent The first main result (Theorem \ref{t1}) is that, under certain regularity conditions on the marginal distribution,   \textit{the second-price auction with the uniformly distributed  random reserve} is a maxmin mechanism across all dominant-strategy  mechanisms for the two-bidder case. Under this mechanism, a random reserve  is drawn from a uniform distribution on $[0,\bar{v}]$ where $\bar{v}$ is the maximum valuation.\\
\indent The randomness in this mechanism hedges against uncertainty over correlation structures. Indeed, the specific random device in this mechanism  exhibits a \textit{full-insurance property}: the expected revenue is the same across all joint distributions consistent with the marginal distribution. To see this, consider a valuation profile $(v_1,v_2)$ in which $v_1>v_2$. Under this mechanism, if the random reserve $r$ is lower than $v_2$, occurring with a probability of $\frac{v_2}{\bar{v}}$, then bidder 1 pays $v_2$; if the random reserve $r$ is between $v_2$ and $v_1$,  then bidder 1 pays $r$. Therefore the revenue from the valuation profile $(v_1,v_2)$ is $v_2\cdot \frac{v_2}{\bar{v}}+\int_{v_1}^{v_2}r\cdot \frac{1}{\bar{v}}dr=\frac{v_1^2+v_2^2}{2\bar{v}}$, which is separable in $v_1$ and $v_2$. This implies the full-insurance property. In addition, the expected revenue is the second moment of the marginal distribution over the maximum valuation. \\
\indent I  show that this mechanism is a maxmin mechanism across all  dominant-strategy  mechanisms under certain regularity conditions by constructing a conceivable joint distribution such that (i) the joint distribution minimizes the expected revenue under the mechanism across all conceivable joint distributions  and (ii) the mechanism maximizes the expected revenue under the joint distribution across all dominant-strategy  mechanisms. I call such a joint distribution a \textit{worst-case correlation structure}. It is straightforward that (i) and (ii) imply that the mechanism is a maxmin mechanism. \\
\indent To construct such a joint distribution, I first reformulate the problem of maximizing the expected revenue across all dominant-strategy mechanisms  as the problem of maximizing the expected ``virtual value'' of the bidder who is allocated the object  subject to that the allocation rule is monotone (a monotonicity constraint associated with dominant-strategy incentive compatibility), where the ``virtual value'' is that the bidder's valuation less information rents that are pinned down by  dominant-strategy incentive compatibility and the binding ex-post participation
constraints of zero-valuation bidders.  This is a straightforward adaption of the well-known revenue equivalence result of \cite{myerson1981optimal}. Importantly,  this simplifies the problem in that  one can now point-wise maximizes the objective, ignoring the monotonicity constriant\footnote{Of course, one need to check that the monotonicity constraint holds in the end.}.    Then such a joint distribution is obtained by  letting the virtual value of the high bidder (the bidder with a higher valuation than that of her opponent) be zero except when the high bidder's valuation is $\bar{v}$. The intuition behind this property is that the auctioneer is \textit{indifferent} between allocating and not allocating the object to the high bidder as long as her valuation is below $\bar{v}$  under the second-price auction with the uniformly distributed  random reserve. \\
\indent To illustrate, consider a special case in which the marginal distribution is an \textit{equal-revenue} distribution\footnote{This distribution is identified as a buyer-optimal signal distribution in a monopoly selling problem by \cite{roesler2017buyer}.}, defined by the property of   a unit-elastic demand: in the monopoly pricing problem, the monopoly's revenue from charging any price in the support of this distribution is the same.  Notably, there is a probability mass on the maximum valuation in an equal-revenue distribution. Under the joint distribution where the high bidder's virtual value is 0 except when her valuation is $\bar{v}$, the two bidders' valuations turn out to be \textit{independent}. Notably, the low bidder's virtual value  also \textit{equals} 0.  The proposed regularity conditions generalize the special case: they guarantee that under the constructed joint distribution, the high bidder's virtual value is 0 and the low bidder's virtual value is \textit{weakly negative}.  Then if the proposed regularity conditions hold,  the second-price auction with the uniformly distributed random reserve maximizes the  expected revenue across all dominant-strategy  mechanisms under the constructed joint distribution. Indeed, if the proposed regularity conditions hold,  any dominant-strategy mechanism,   in which 1) the ex-post participation constraints are binding for zero-valuation bidders and 2) the object is allocated with probability one to the high bidder with the valuation of $\bar{v}$ and the object is never allocated to the low bidder,  maximizes the expected revenue across all dominant-strategy mechanisms under the constructed joint distribution.  \\
\indent Notably, the proposed regularity conditions contain a probability mass condition on the maximum valuation. That is, the result requires that the marginal distribution have an atom on the maximum valuation and that the size of the atom be bounded from below.  Indeed,  these conditions capture many heavy-tailed distributions\footnote{I present a detailed discussion of these conditions in Section \ref{s51}.}, which are  observed in many real-world auctions. For example, according to  \cite{arnosti2016adverse},  it has been observed that  a huge fraction
of the total value comes from a small number of very valuable impressions  in online advertising.  In addition, according to \cite{ibragimov2010optimal},   very diverse private valuations have been observed in markets for cultural and sport
events as well as in those for antiques and collectibles
and online auctions and marketplaces such as eBay
and StubHub.  Many papers have studied mechanism design problems when the distribution of values exhibits a heavy tail (e.g., \cite{arnosti2016adverse} and \cite{ibragimov2010optimal}). \\
\indent I extend the analysis to the case of general number of bidders $(N\ge 3)$.  For tractability, I restrict attention  to a subclass of dominant-strategy  mechanisms in which a bidder whose bid is not the highest is never allocated. A mechanism in this subclass is referred to as a \textit{standard}\footnote{The terminology of ''standard'' comes from \cite{bergemann2019revenue} who define
standard mechanisms in a similar manner. \cite{he2022correlation} also adopts this terminology.} dominant-strategy  mechanism. The second main result (Theorem \ref{t2}) is that, under certain regularity conditions on the marginal distribution,  \textit{the second-price auction with the $\bar{v}-$scaled $Beta (\frac{1}{N-1},1)$ distributed random reserve} is a maxmin mechanism across  standard dominant-strategy  mechanisms.  Under this mechanism, the cumulative distribution function of the random reserve $r$ is $G(r)=(\frac{r}{\bar{v}})^{\frac{1}{N-1}}$ for $r\in[0,\bar{v}]$. Following a methodology similar to  the two-bidder case, I  show  this result by constructing a worst-case correlation structure. \\
\indent This mechanism embodies, albeit not the full-insurance property,  a good hedging property: it yields the same expected revenue across a range of correlation structures. Precisely, as I will show,  the expected revenue is the same for any conceivable joint distribution whose support lies in  the set of  valuation profiles in which either all bidders have the same valuations or there is a unique highest bidder and the other bidders have the same valuations. Indeed, the constructed worst-case correlation structure  has such a support. Intuitively, given the restriction to standard mechanisms, only the highest bidders are possible to generate positive revenue to the auctioneer. Thus, the other bidders' valuations except for the highest one are  ``wasted''.  To reduce the expected  revenue as much as possible while maintaining the consistency with the marginal distribution, a worst-case correlation structure maximizes the waste by increasing the other bidders' valuations as much as possible until all the other bidders' valuations are the same. Then similar to the two-bidder case, the worst-case  correlation structure is obtained by requiring the highest bidder's virtual value be zero except when the highest bidder's valuation is $\bar{v}$.  Here, the proposed regularity conditions on the marginal distribution guarantee that the construction is feasible. As I focus on standard dominant-strategy mechanisms,  the bidders whose valuations are not the highest do not contribute to the expected revenue and therefore their virtual values do not matter. Then it is straightforward that the second-price auction with the  $\bar{v}-$scaled $Beta (\frac{1}{N-1},1)$ distributed random reserve maximizes the expected revenue across  standard dominant-strategy  mechanisms under the constructed joint distribution.   \\
\indent Moreover, I show that  the second-price auction with the $\bar{v}-$scaled $Beta (\frac{1}{N-1},1)$ distributed random reserve is asymptotically optimal across all dominant-strategy mechanisms as the number of the bidders goes to infinity, regardless of the marginal distribution (Remark \ref{r4}). To establish this, I show that the revenue guarantee of this mechanism converges to the expectation of a generic bidder's valuation.  Indeed, the expectation of a generic bidder's valuation is an upper bound of the revenue guarantee for any dominant-strategy mechanism, as it is possible that the correlation structure is the maximally positively correlated distribution (that is, all bidders have the same valuations for any valuation profile in the support), under which the expectation of a generic bidder's valuation is the most surplus  that the auctioneer can extract. \\
\indent The first two main results both require the probability mass on the maximum valuation to be bounded from below. The third main result (Theorem \ref{t5}) characterizes \textit{the second-price auction with the  $s^*\footnote{$s^*$ is characterized by the known marginal distribution, details of which are given in Section \ref{ae}.}-$scaled   $Beta (\frac{1}{N-1},1)$ distributed random reserve} as a  maxmin mechanism across standard dominant-strategy mechanisms if the probability mass condition does not hold. Under this mechanism, the cumulative distribution function of the random reserve $r$ is $G_{s^*}(r)=(\frac{r}{s^*})^{\frac{1}{N-1}}$ with support $[0,s^*]$ where $s^*\in (0,\bar{v})$. Notably, the highest bidder will be fully allocated the object provided that her valuation is higher than $s^*$. \\
\indent Remarkably,   a second-price auction (albeit with some random reserve), which is simple and widely adopted in practice,  arises as a robustly optimal mechanism across some general class of mechanisms. Importantly, this is true for a wide range of marginal distributions.     Therefore, the main  results provide a positive explanation why the second-price auction is prevalent in the real world: it maximizes the worst-case expected revenue for a wide range of marginal distributions. Furthermore, the explanation is particularly convincing for the two-bidder case: the robust optimality is established across \textit{all} dominant-strategy mechanisms. \\
\indent In addition to the main results, I propose a family of second-price auctions with $t-$scaled  $Beta (\frac{1}{N-1},1)$ distributed random reserves where $t\in (0,\bar{v})$. I identify a non-trivial lower bound of the revenue guarantee for each auction in this family. For any given marginal distribution,    I am able to find an auction from this family  that has a  strictly higher revenue guarantee than  that of any  posted-price mechanism and any  second-price auction with a non-negative deterministic reserve (Theorem \ref{t3} and Theorem \ref{t4}). \\
\indent The remainder of the introduction discusses related literature. 
Section \ref{s3} presents the model. Section \ref{s4} illustrates the methodology and conducts preliminary analysis.    Section \ref{s5} characterizes the main  results. Section \ref{s6}  proposes a family of auctions and studies their performance in terms of the revenue guarantee. Section \ref{s7} is a conclusion.  Omitted  proofs are in  Appendix \ref{aa}, \ref{ab} and \ref{ac}. Appendix \ref{ad} contains the result for the essential necessity of the  regularity conditions.

\subsection{Related Literature}\label{s2}
The closest related paper is \cite{he2022correlation}, who study the design of auctions within the correlation-robust framework. They show, among others, that  a second-price auction with a random reserve is a maxmin mechanism across standard dominant-strategy mechanisms under certain conditions on the marginal distribution. Methodologically, both papers use duality theory to proceed the analysis.  The main differences can be summarized as follows.   My setting is more general than theirs in that I allow the marginal distribution to have a probability mass on the maximum valuation\footnote{They assume continuous distributions and do not allow for a mass point on the maximum valuation.}. More importantly,  I obtain a new and strong result for the two-bidder case: I establish  that,  under certain regularity conditions (different from theirs),  my proposed mechanism is a maxmin mechanism across \textit{all} dominant-strategy mechanisms.  In addition,  I establish the main results under weaker conditions on the marginal distribution\footnote{For valuations below the maximum valuation, both papers assume that the marginal distribution admits a density $f$. For the main results, I only require that $x^2f(x)$ be non-decreasing in $x$  instead of that $xf(x)$ be non-decreasing in $x$, required in their paper.}.  \\
\indent This paper is  closely related to \cite{che2020distributionally}, \cite{koccyiugit2020distributionally} and \cite{zhang2021robust}. The first two papers both  consider a model of auction design in which the auctioneer only knows the expectation of each bidder's valuation. Specifically, \cite{che2020distributionally} shows that a second-price auction with  an optimal random reserve is  a maxmin mechanism within
a class of competitive mechanisms\footnote{The class of dominant-strategy mechanisms is not a subset of the class of competitive mechanisms, and vice versa.}; \cite{koccyiugit2020distributionally} characterize a  maxmin mechanism across highest-bidder lotteries\footnote{This is the same as standard dominant-strategy  mechanisms. } for the case where the known expectations are the same across bidders. Similarly, my paper also considers some general class of mechanisms. The main difference is that I assume that the auctioneer knows exactly the marginal distribution. That is, I assume that the auctioneer knows more and therefore the revenue guarantee in my setting is an upper bound of theirs. \cite{zhang2021robust} considers a model of bilateral trade in which the profit-maximizing intermediary only knows the expectations of each trader's valuation. He characterizes maxmin trading mechanisms across all dominant-strategy mechanisms. The maxmin trading mechanism features fixed-commission fee, uniformly random spread and midpoint transaction price in the symmetric case. In contrast, this paper considers a model of auction design and assumes that the auctioneer knows exactly the marginal distribution. Like that paper, I consider all dominant-strategy mechanisms for the two-bidder case. In addition,   this paper employs a similar methodology to proceed the analysis. More specifically, both papers use properties of ``virtual value'' to construct  worst-case distributions.    \\
\indent This paper is also closely related to \cite{bei2019correlation}, who study the design of auctions within  the correlation-robust framework with a focus on \textit{simple} mechanisms. They show, among others, that the revenue guarantee of  the sequential posted-price mechanism  is at least $\frac{1}{2\ln{4}+2}$ times the revenue guarantee of the optimal dominant-strategy mechanism.  \\
\indent This paper is related to \cite{bose2006optimal} who study the design of auctions assuming that the bidders' valuations are independently distributed but there may be ambiguity about the marginal distribution of a generic bidder's valuation. In contrast, my paper assumes that the marginal distribution of a generic bidder's valuation is known but the correlation structure of bidders' valuations is unknown. Because of the different framework,  the methodology of this paper differs significantly from that one. They show,  among others, that an auction that ``fully insures'' the auctioneer is a maxmin mechanism when the auctioneer does not know the marginal distribution but the bidders know it. Similarly, in my framework,  the first main result (Theorem \ref{t1}) characterizes a maxmin mechanism exhibiting  a full-insurance property. However, the notion of full-insurance is different from theirs. While full-insurance requires that  the same \textit{expected revenue} be obtained across all conceivable distributions in this paper, it  requires that  the same \textit{ex-post revenue} be obtained across all valuation profiles in that one. \\
\indent Broadly, this paper joins the robust mechanism design literature (\cite{bergemann2005robust}). There are other papers  searching optimal solutions in the worst case over the space of parameters (e.g., \cite{carroll2016robust},  \cite{garrett2014robustness}, \cite{bergemann2011robust}, \cite{carroll2017robustness},\cite{giannakopoulos2020robust},\cite{chen2019distribution}).
 \cite{bergemann2016informationally}, \cite{du2018robust} and \cite{brooks2020optimal}   consider a model of auction design with  common values. They assume that 
 bidders’ valuations for the object are drawn from a commonly known prior, but they may have
arbitrary information (high-order beliefs) about
the prior distribution unknown to the seller. An auction’s
performance is measured by the worst expected revenue across a class of incomplete information
correlated equilibria termed Bayes correlated equilibria
(BCE) in \cite{bergemann2013robust}. In this paper, I completely ignore the
beliefs of  bidders by focusing on the dominant-strategy  mechanisms.  This assumption is more appropriate for situations in which one not say much about  bidders' beliefs, as dominant-strategy mechanisms are robust to misspecification of bidders'  beliefs. 
\section{Preliminaries}\label{s3}
\subsection{Notation}
I introduce the following technical notations. First, all spaces considered are polish spaces; I endow them with their Borel $\sigma-$algebra. Second, product spaces are endowed with product $\sigma-$algebra. Third, I use $\Delta (X)$ to denote the set of all probability measures over $X$. 
\subsection{Environment}
I consider an environment where a  single indivisible object is sold to $N\ge 2$ risk-neutral bidders. I denote by $I = \{1, 2, . . . , N\}$ the set of bidders.  Each bidder $i$ has private information about her
valuation for the object, which is modeled as a random variable $v_i$ with cumulative distribution
function $F_i$\footnote{As will be discussed later, I allow distributions to have a probability mass on the maximum valuation. Furthermore,  all results (with slight modifications) hold in discrete environments.}.   Throughout the paper, I focus on symmetric environment, i.e., $F_i=F_j=F$\footnote{With slight abuse of notation, I also use $F$ to denote the probability measure consistent with the distribution $F$. } for any $i,j \in I $. I denote  the support of  $F_i$ by $V_i$.  I assume that $V_i = [0,\bar{v}]$ for some $\bar{v}>0$.  The joint support of all $F_i$
is  $V= \times_{i=1}^N V_i=[0,\bar{v}]^N$ with a typical valuation profile $v$. I denote bidder $i$'s opponents' valuation profiles by $v_{-i}$, i.e., $v_{-i}\in V_{-i}=\times_{j\neq i}V_j$. \\
\indent The valuation profile $v$ is drawn from a joint distribution $\mathcal{P}$,
which  may have an arbitrary correlation structure. The auctioneer only knows the
marginal distribution $F$ of each bidder's valuation but does not know how these bidders' valuations are
correlated.   To the auctioneer, any joint distribution is \textit{conceivable}
as long as the joint distribution is consistent with the marginal distribution. I denote by
\[\Pi(F) = \{
\pi\footnote{With slight abuse of notation, I also use $\pi$ to denote the probability density of the probability measure $\pi$ if the probability density exists.} \in \Delta V : \forall i \in I, \forall A_i \subset  V_i
, \pi(A_i \times V_{-i}) = F(A_i)\}\]
the collection of conceivable joint distributions. 
\subsection{Marginal Distribution}\label{s33}
I assume that the marginal distribution $F$ admits a probability density function  $f(x)$ for any $x\in [0,\bar{v})$. I allow $F$ to have a probability mass on $\bar{v}$, the size of which is  denoted by $Pr(\bar{v})$. \footnote{\cite{he2022correlation} restrict attention to marginal distributions that admits a positive density function everywhere, whereas this paper allows a probability mass on the maximum valuation. That is, the assumption in this paper is more general.} Importantly, as will be seen, the first two main results (Theorem \ref{t1} and \ref{t2}) require that  the marginal distribution have an atom on $\bar{v}$ and that the size of the atom be  bounded from below.   Indeed, distributions with a probability mass on the maximum valuation are familiar in many information design and  robust mechanism design environments. \cite{roesler2017buyer} analyze a model where the buyer has a given value distribution and designs a signal structure to learn about her
valuation. After observing the signal structure, the seller makes a take-it-or-leave-it offer. They show that an equal-revenue distribution is the buyer-optimal signal distribution. In a closely related work, \cite{condorelli2020information} analyze a model where the buyer can choose the probability distribution
of her valuation for the good. After observing the buyer’s choice of the distribution, the seller  makes a take-it-or-leave-it offer. They  show that an equal-revenue distribution is the buyer-optimal value distribution. Besides,  \cite{bergemann2008pricing} consider a minimax regret design problem of selling a single object to a single buyer and find that an equal-revenue distribution is a worst-case distribution;  \cite{zhang2022robust} considers a  robust public good mechanism design problem and finds that the  worst-case marginal distribution has a probability  mass on the maximum valuation.\\
\indent Moreover, distributions with a probability mass on the maximum valuation admits an ''approximating'' interpretation as follows. Consider a  marginal distribution $\hat{F}$ with a non-negative density function $\hat{f}$ everywhere on $[0,\infty)$. In the real world, it is reasonable to assume that bidders' valuations are bounded as the total wealth, which is physically impossible to be infinite, is an upper bound of bidders' valuations. Then  the marginal distribution  $F$ with bounded support $[0,\bar{v}]$ is  generated by   truncating  $\hat{F}$ on $\bar{v}$ as follows: $f(x)=\hat{f}(x)$ for $x<\bar{v}$ and $Pr(\bar{v})=1-\hat{F}(\bar{v})$. When $\bar{v}$ is large, $F$ is a natural approximation of $\hat{F}$. 
\subsection{(Standard) Dominant-strategy Mechanisms}
\indent I focus on dominant-strategy mechanisms. The revelation principle holds and it is without loss of generality to restrict attention to direct mechanisms.   A direct mechanism  $(q,t)$ is defined as an allocation rule $q : V \to [0, 1]^N$ and a payment function $t : V \to \mathbb{R}^N$. With slight abuse of notation, each bidder submits a sealed bid $v_i\in V_i$ to the auctioneer. Upon receiving the bids profile $v=(v_1, v_2, \cdots, v_N)$, the allocation probabilities are $q(v)=(q_1(v), q_2(v),\cdots, q_N(v))$ and the payments are $t(v)=(t_1(v), t_2(v),\cdots, t_N(v))$ where $q(v)\ge 0 $ and $\sum_i q_i(v)\le 1 $ for all $v\in V$.   A direct mechanism is a dominant-strategy mechanism  if for all $i\in I$, all $v\in V$, and all $v_i'\in V_i$, 
\[v_iq_i(v)-t_i(v)\ge v_iq_i(v_i',v_{-i})-t_i(v_i',v_{-i}), \]
\[v_iq_i(v)-t_i(v)\ge 0. \]
The set of all such  mechanisms is denoted by $\mathcal{D}$.  I say a direct mechanism $(q,t)$ is \textit{standard} if for any $v \in V$ and $i \in I$ such that $v_i < \max_{j\in I} v_j$ , the allocation to the bidder $i$ is  $q_i(v) = 0$. That is, only the highest bidders are possible to be allocated in a standard  mechanism. 
 The set of all standard dominant-strategy  mechanisms is denoted by $\mathcal{E}$.
 \subsection{Objective Function}
\indent I am interested in the auctioneer’s expected revenue in the dominant-strategy equilibrium in which each bidder truthfully reports her valuation of the
object. Then the expected revenue of a dominant-strategy  mechanism $(q,t)$ when the joint distribution is $\pi$ is $U((q,t),\pi)=\int_{v\in V}\sum_{i=1}^Nt_i(v) d\pi(v)$. The auctioneer evaluates a mechanism $(q,t)$ by its worst-case expected revenue, referred to as the \textit{revenue guarantee} of the mechanism $(q,t)$,  over all conceivable joint distributions. Formally, the mechanism $(q,t)$'s revenue guarantee is $REG((q,t))=\inf_{\pi\in \Pi(F)}U((q,t),\pi)$.  The auctioneer's goal is to find a \textit{maxmin mechanism} from either $\mathcal{D}$  or $\mathcal{E}$  with the maximal revenue guarantee\footnote{As will be seen later, the first main result (Theorem \ref{t1}) characterizes a maxmin mechanism from $\mathcal{D}$, and either of the other main results (Theorem \ref{t2} and Theorem \ref{t5}) characterizes a maxmin mechanism from $\mathcal{E}$.}. Formally, the auctioneer  solves \[
\sup_{(q,t)\in \mathcal{D}(\text{or} \mathcal{E})}REG((q,t)).\tag{MRG}\label{mrg}\]
\section{Methodology and Preliminary Analysis}\label{s4}
The maxmin optimization problem \eqref{mrg}  can be interpreted as a two-player sequential zero-sum
game. The two players are the auctioneer and adversarial nature. The auctioneer first chooses
a mechanism $(q,t)\in \mathcal{D}$ (or $(q,t)\in \mathcal{E}$). After observing the auctioneer’s choice of
the mechanism, adversarial nature chooses a conceivable joint distribution  $\pi\in \Pi(F)$. The auctioneer’s
payoff is $U((q,t),\pi)$, and adversarial nature’s payoff is $-U((q,t),\pi)$. One can also consider the simultaneous-move version of this zero-sum game, whose Nash equilibrium is indeed a \textit{saddle point} of the payoff functional $U$, i.e., for any $(q,t)\in \mathcal{D}$ (or $(q,t)\in \mathcal{E}$) and any $\pi\in \Pi(F)$, \[U((q^*,t^*), \pi) \ge U((q^*,t^*),\pi^*) \ge U((q,t),\pi^*). \]
The first inequality  says the joint distribution $\pi^*$ minimizes the expected revenue   under the mechanism $(q^*,t^*)$, and the second inequality implies that,  under the joint distribution $\pi^*$, the other dominant-strategy  mechanisms cannot attain a strictly higher expected revenue. Hence, the auctioneer’s
equilibrium strategy in the simultaneous-move version of this zero-sum game, $(q^*,t^*)$, is a maxmin mechanism.   $\pi^*$ is referred to as a \textit{worst-case correlation structure}.  I will construct a saddle point for each of  the main results. 
\begin{proposition}[Revenue Equivalence]\label{p1}
When searching for a maxmin mechanism, it is without loss to restrict attention to mechanisms satisfying the following properties: 1) $q_i(\cdot,v_{-i})$ is non-decreasing in $v_i$ for all $v_{-i}$ and 2)  $t_i(v_i,v_{-i})=v_iq_i(v_i,v_{-i})-\int_0^{v_i}q_i(x,v_{-i})dx$.
\end{proposition} 
\begin{proof}
The proof is in  Appendix \ref{aa}.
\end{proof}
Proposition \ref{p1} simplifies the analysis by establishing two properties of a maxmin mechanism. The first property says the allocation rule is \textit{monotone}, and the second property says that the payment rule can be characterized by the allocation rule and that the ex-post participation constraints are binding for zero-valuation bidders. This is standard in the mechanism design literature (e.g., \cite{myerson1981optimal}).\\
\indent Moreover, Proposition \ref{p1} allows me to obtain a virtual representation of the expected revenue, which is essential for my analysis. Precisely, consider the problem that  fixing an arbitrary joint distribution $\pi$, the auctioneer  designs an optimal mechanism $(q,t)$. For exposition, I assume that $\pi$ admits a positive density function\footnote{The virtual representation can be similarly derived for joint distributions in which there is a probability mass on $\underbrace{(1,\cdots,1)}_{N}$.}.   The density function of $v_i$ conditional on $v_{-i}$ is denoted by $\pi_i(v_i| v_{-i})$, and the cumulative  distribution function of $v_i$ conditional on $v_{-i}$ is denoted by  $\Pi_i(v_i|v_{-i})$. Then an direct implication of Proposition \ref{p1} is that the expected revenue of $(q,t)$ under the joint distribution $\pi$ is \[E[\sum_{i=1}^Nt_i(v)]=E[\sum_{i=1}^Nq_i(v)\phi_i(v)],\]
where $\phi_i(v)= 
v_i-\frac{1-\Pi_i(v_i|v_{-i})}{\pi_i(v_i|v_{-i})}$ is the \textit{virtual value} of bidder $i$ when the valuation profile is $v$. Thus the problem of designing an optimal mechanism given a joint distribution is equivalent to maximizing the \textit{expected total virtual surplus}, which refers to  the expected sum of the allocation times the  virtual value,  subject to that the allocation rule is monotone.
\section{Main Results}\label{s5}
In Section \ref{s51}, I characterize a maxmin mechanism across all dominant-strategy  mechanisms under certain regularity conditions for the two-bidder case. In Section \ref{s52}, I characterize a maxmin mechanism across standard dominant-strategy  mechanisms under certain regularity conditions for the $N-$bidder ($N\ge 3$) case.
In Section \ref{ae}, I characterize a maxmin mechanism across standard dominant-strategy  mechanisms condition for the $N-$bidder ($N\ge 2$) case when the probability mass condition  (part of the regularity conditions) fails.
\subsection{Two Bidders}\label{s51}
I first define a mechanism  and a joint distribution. Then I define regularity conditions under which the mechanism and the joint distribution form a saddle point (Theorem \ref{t1}). Then I illustrate Theorem \ref{t1}. Finally, I give a discussion of the regularity conditions. \\
\indent \textit{The second-price auction with the uniformly distributed random reserve} is defined as follows. The auctioneer first draws a random reserve $r$ from the uniform distribution with support $[0,\bar{v}]$. Then the two bidders bid simultaneously.  The high bidder (the bidder with a higher bid than that of her opponent)  wins the object if her bid is also higher than $r$, and she pays the maximum of $r$ and her opponent's bid; the low bidder loses the auction and pays nothing. In case of ties, each bidder wins the object with a half probability if the bid is higher than $r$, and the winner pays the bid. \\
\indent  Equivalently, it can be defined by $(q^*,t^*)$ as follows. If $v_1>v_2$, then $q^*_1(v_1,v_2)=\frac{v_1}{\bar{v}}, q^*_2(v_1,v_2)=0$ and $t^*_1(v_1,v_2)=\frac{v_1^2+v_2^2}{2\bar{v}}, t^*_2(v_1,v_2)=0$; if $v_1<v_2$, then $q^*_1(v_1,v_2)=0, q^*_2(v_1,v_2)=\frac{v_2}{\bar{v}}$ and $t^*_1(v_1,v_2)=0, t^*_2(v_1,v_2)=\frac{v_1^2+v_2^2}{2\bar{v}}$; if $v_1=v_2=x$, then $q^*_1(v_1,v_2)=q^*_2(v_1,v_2)=\frac{x}{2\bar{v}}$ and $t^*_1(v_1,v_2)=t^*_2(v_1,v_2)=\frac{x^2}{2\bar{v}}$.\\
\indent The joint distribution $\pi^*$ is defined as follows\footnote{Here $\pi^*(v_1,v_2)$ denotes the density of the valuation profile $(v_1,v_2)$ whenever the density exists and  $Pr^*(v_1,v_2)$ denotes the probability mass of the valuation profile $(v_1,v_2)$ whenever there is some probability mass on $(v_1,v_2)$. The marginal distributions that  the result covers have a probability mass on the maximum valuation $\bar{v}$. In the joint distribution $\pi^*$, there is (non-negative) probability mass on the point $(\bar{v},\bar{v})$.}. 
\[\pi^*(v_1,v_2)= \pi^*(v_2,v_1)= \left\{
\begin{array}{lll}
f(0)      &      & \\
\qquad \qquad\qquad \qquad \qquad\qquad \text{if $v_1=v_2=0$;}      &      & \\
0      &      & \\
\qquad \qquad\qquad \qquad \qquad\qquad \text{if $v_1>v_2=0  $;}      &      & \\
\frac{1}{v_1^2}(v_2f(v_2)-\frac{\int_{0}^{v_2}x^2f(x)dx}{v_2^2})       &      & \\
\qquad \qquad\qquad \qquad \qquad\qquad \text{if $\bar{v}>v_1 \ge v_2>0$;}      &      & \\
\frac{1}{\bar{v}}(v_2f(v_2)-\frac{\int_{0}^{v_2}x^2f(x)dx}{v_2^2})     &      & \\
\qquad \qquad\qquad \qquad \qquad\qquad \text{if $ \bar{v}=v_1>v_2>0$.}      &      & 
\end{array} \right. \]
\[Pr^*(\bar{v},\bar{v})=Pr(\bar{v})-\frac{\int_{x\in(0,\bar{v})}x^2f(x)dx}{\bar{v}^2}. \]
\indent \textit{The two-bidder robust regularity conditions} are defined as follows:  $x^2f(x)$ is non-decreasing for $x\in(0,\bar{v})$ and $Pr(\bar{v}) \ge \frac{\int_{x\in (0,\bar{v})}x^2f(x)dx}{\bar{v}^2}$. I refer to the second part of the conditions as the probability mass condition.
\begin{remark}
\normalfont Note that the probability  mass condition will vanish as $\bar{v}\rightarrow \infty$ if $\int_{x\in (0,\bar{v})}x^2f(x)dx$ is of order $\bar{v}^\gamma$ with $\gamma<2$. 
\end{remark}
\begin{theorem}\label{t1}
For the two-bidder case, the second-price auction with the  uniformly distributed random reserve is a maxmin mechanism across all dominant-strategy  mechanisms if the two-bidder robust regularity conditions hold.  The revenue guarantee is $\frac{E[X^2]}{\bar{v}}$.\footnote{The distribution of $X$ is $F$. } The joint distribution $\pi^*$ is a worst-case correlation structure.
\end{theorem}
Now I illustrate Theorem \ref{t1}. I start with the illustration of the mechanism. 
\begin{definition}
\normalfont I say a dominant-strategy mechanism $(q,t)$ exhibits the \textit{full-insurance property } if the expected revenue of $(q,t)$ is the same across all conceivable joint distributions. 
\end{definition}
\begin{proposition}\label{p2}
For the two-bidder case, the second-price auction with the  uniformly distributed random reserve exhibits the full-insurance property.
\end{proposition}
\begin{proof}
Note that under the second-price auction with the uniformly distributed random reserve,  the total revenue from a valuation profile $(v_1,v_2)$ is  $t^*(v_1,v_2)=t_1^*(v_1,v_2)+t_2^*(v_1,v_2)=\frac{v_1^2+v_2^2}{2\bar{v}}$. Then fix  any conceivable joint distribution $\pi$, the expected revenue   is 
\[
    \begin{split}
        \int_{[0,\bar{v}]^2}t^*(v_1,v_2)d\pi(v_1,v_2)&=
    \int_{[0,\bar{v}]^2}\frac{v_1^2+v_2^2}{2\bar{v}}d\pi(v_1,v_2)\\
    &=\int_{[0,\bar{v}]^2}\frac{v_1^2}{2\bar{v}}d\pi(v_1,v_2)+\int_{[0,\bar{v}]^2}\frac{v_2^2}{2\bar{v}}d\pi(v_1,v_2)\\
    &=\int_{[0,\bar{v}]}\frac{v_1^2}{2\bar{v}}dF(v_1)+\int_{[0,\bar{v}]}\frac{v_2^2}{2\bar{v}}dF(v_2)\\
    &=\frac{E[X^2]}{\bar{v}}.
    \end{split}\]
\end{proof}
\indent The joint distribution $\pi^*$ is obtained by a condition requiring  that the high bidder's virtual value be 0 except when her valuation is $\bar{v}$. Formally, \[
\phi^*_i(v_i,v_j)=0 \quad \text{if}\quad v_j\le v_i<\bar{v}.\tag{1}\label{1}\]
The property \eqref{1} is motivated by a property of the second-price auction with the uniformly distributed random reserve:  the auctioneer is indifferent between allocating and not allocating the object to the high bidder  as long as her valuation is not $\bar{v}$. \\
\indent Furthermore, I impose a condition on the constructed joint distribution $\pi^*$ that the virtual value of the low bidder is weakly smaller than that of the high bidder. Formally, 
 \[\phi^*_j(v_i,v_j)\le \phi_i^*(v_i,v_j) \quad \text{if}\quad v_j\le v_i.\tag{2}\label{2}\]
 The property \eqref{2} is motivated by another property of the second-price auction with the uniformly distributed random reserve: the low bidder is never allocated the object. \\
 \indent Indeed, if  the property \eqref{1} and the property   \eqref{2} hold for a joint distribution, it is straightforward that the second-price auction with the  uniformly distributed random reserve maximizes the expected revenue across all dominant-strategy mechanisms given this joint distribution. The two-bidder robust regularity conditions, as I will show, guarantee that the property \eqref{1} and the property \eqref{2} hold for the constructed joint distribution $\pi^*$. In summary, I obtain a   virtual value matrix  for $\pi^*$  as follows if the two-bidder robust regularity conditions hold.
\begin{equation*}
\begin{pmatrix}
(0,0)_{0,0} & (0,0)_{0,>0} & \cdots & \cdots & (0,0)_{0,<\bar{v}} & (0,0)_{0,\bar{v}}\\
(0,0)_{>0,0} & (0,0)_{v_1=v_2>0} & (-,0)_{v_1<v_2}  & \cdots& (-,0)_{v_1<v_2<1}& (\le ,+)_{<\bar{v},\bar{v}} \\
\vdots  & (0,-)_{v_1>v_2}  & \ddots & \vdots& \vdots& \vdots  \\
\vdots & \vdots & \cdots & \ddots & \vdots & \vdots\\
(0,0)_{<\bar{v},0} & (0,-)_{<\bar{v},>0} & \cdots & \cdots & (0,0)_{v_1=v_2<\bar{v}} & (\le,+)_{<\bar{v},\bar{v}}\\
(0,0)_{\bar{v},0} & (+,\le)_{\bar{v},>0} & \cdots &\cdots & (+,\le)_{\bar{v},<\bar{v}} & (+,+)_{\bar{v},\bar{v}}
\end{pmatrix}
\end{equation*}
Here ``0'' in the bracket means zero virtual value, ``$-$'' means a non-positive virtual value, ``+'' means a non-negative virtual value, ``$\le$'' means the virtual value of the bidder is weakly smaller than that of her opponent. The subscript denotes the corresponding valuation profile.
\begin{remark}
\normalfont The probability mass condition arises because the property \eqref{1} requires that the conditional distribution of the high bidder's valuation be an equal-revenue distribution, which has an atom on the maximum valuation $\bar{v}$.
\end{remark}
 \begin{proposition}\label{p3}
 If the two-bidder robust regularity conditions hold, the second-price auction with  the uniformly distributed random reserve maximizes the expected revenue across all dominant-strategy mechanism under the joint distribution $\pi^*$.
 \end{proposition}
 \begin{proof}
 The  proof is in Appendix \ref{ab} which presents details about the construction of $\pi^*$.
 \end{proof}
 \indent Theorem \ref{1} follows immediately from Proposition \ref{p2} and Proposition \ref{p3}.\\
\indent There is an important special case where Theorem \ref{t1} applies:  the equal-revenue distribution. Recall that the equal-revenue distribution is familiar in the information design literature and robust mechanism design literature (e.g., \cite{roesler2017buyer}, \cite{bergemann2008pricing}, \cite{du2018robust}, etc). 
\begin{corollary}\label{c2}
For the two-bidder case, if the marginal distribution is an equal-revenue distribution with $\alpha \in (0,\bar{v})$: \[
F(x)=
\left\{
\begin{array}{rcl}
1-\frac{\alpha}{x}     &      & {\normalfont \text{if}\quad\alpha \le x< \bar{v};}\\
1      &      & {\normalfont \text{if}\quad x=\bar{v},}
\end{array} \right.\]
then the second-price auction with the uniformly distributed random reserve is a maxmin mechanism across all dominant-strategy mechanism. The revenue guarantee is $2\alpha-\frac{\alpha^2}{\bar{v}}$.  The independent equal-revenue distribution\footnote{That is, the marginal distribution of each bidder's valuation is the known equal-revenue distribution; bidders' valuations are independently distributed. } is a worst-case correlation structure. 
\end{corollary}
\begin{proof}
It is straightforward to show that an equal-revenue distribution satisfies the two-bidder robust regularity conditions. Then Theorem \ref{t1} implies Corollary \ref{c2}.
\end{proof}
\indent There are many other distributions satisfying the two-bidder robust regularity conditions. I now provide some examples.
\begin{example}
\normalfont Any (truncated) Pareto distribution with $\alpha\in (0,\bar{v}), \beta \in (0,1)$: \[
F(x)=
\left\{
\begin{array}{rcl}
1-\frac{\alpha^\beta}{x^\beta}     &      & {\normalfont \text{if}\quad\alpha\le x < \bar{v};}\\
1    &      & {\normalfont \text{if}\quad x=\bar{v}.}
\end{array} \right.\]
\end{example}
To see this, note that $x^2f(x)=\alpha^\beta\beta x^{1-\beta}$ is non-decreasing when $\beta\in (0,1)$. For the probability mass condition,  note that  $Pr(\bar{v})=(\frac{\alpha}{\bar{v}})^\beta \ge (\frac{\alpha}{\bar{v}})^\beta \frac{\beta}{2-\beta}[1-(\frac{\alpha}{\bar{v}})^{2-\beta}]=\frac{\int_{(0,\bar{v})}x^2f(x)dx}{\bar{v}^2}$ when $\beta\in (0,1)$.
\begin{example}
\normalfont A combination of an uniform distribution on $[0,\bar{v})$ and a  probability  mass on $\bar{v}$ with  $Pr(\bar{v})\ge \frac{1}{4}$.
\end{example}
To see this, note that the first part of the conditions holds trivially because it is uniformly distributed on $[0,\bar{v})$. For the probability mass condition,  note that  $Pr(\bar{v})\ge \frac{\int_{(0,\bar{v})}x^2\cdot \frac{1-Pr(\bar{v})}{\bar{v}}dx}{\bar{v}^2}=\frac{\int_{(0,\bar{v})}x^2f(x)dx}{\bar{v}^2}$ when $Pr(\bar{v})\ge \frac{1}{4}$.\\\\
\indent As a final topic of this section, I discuss the two-bidder robust regularity conditions. Using the approximation interpretation in Section \ref{s33}, $F$ is obtained via a truncation of $\hat{F}$ on $\bar{v}$. Now consider the following regularity conditions  for $\hat{F}$ : $x^2\hat{f}(x)$ is non-decreasing on $[0,\infty)$ and $\frac{\int_0^s x^2\hat{f}(x)dx}{s^2}\rightarrow 0$ as $s\rightarrow \infty$. These conditions  imply that the two-bidder robust regularity conditions hold for any $\bar{v}>0$.\footnote{Indeed,  that $x^2\hat{f}(x)$ is non-decreasing implies that the function $K(s):=\hat{F}(s)+\frac{\int_0^s x^2\hat{f}(x)dx}{s^2}$ is non-decreasing. To see this, note that $K'(s)=2[\hat{f}(s)-\frac{\int_0^sx^2\hat{f}(x)dx}{s^3}]\ge 0$, shown in \eqref{eq20}. Then $\hat{F}(\bar{v})+\frac{\int_0^{\bar{v}} x^2\hat{f}(x)dx}{\bar{v}^2}\le 1$ for any $\bar{v}>0$ if  $\frac{\int_0^s x^2\hat{f}(x)dx}{s^2}\rightarrow 0$ as $s\rightarrow \infty$.} It is straightforward to verify that these conditions hold for many heavy-tailed distributions  including a family of power law distributions\footnote{The power law distribution is given by $F(x)=1-\frac{\alpha^{\beta}}{x^\beta}$ with $\alpha>0$ for $x\in [\alpha, \infty)$. The parameter $\beta$ determines the weight of the tail. These conditions hold for any power law distribution with $\beta\in (0,1]$.  }, Cauchy distributions\footnote{The density function of the Cauchy distribution is given by $f(x)=\frac{2b}{\pi (b^2+x^2)}$ with $b>0$ for $x\in [0,\infty)$. It is straightforward to show that these conditions hold for Cauchy distributions with any  $b>0$. }, log-Cauchy distributions\footnote{The density function of the log-Cauchy distribution is given by $f(x)=\frac{1}{\pi x}[\frac{\sigma}{(\ln{x}-\mu)^2+\sigma^2}]$ with $\sigma>0$ for $x\in (0,\infty)$. It is straightforward to show that these conditions hold for log-Cauchy distributions with any  $\sigma \ge 1$ and any real number $\mu$.}, L\'evy distributions\footnote{The density function of the L\'evy distribution is given by $f(x)=\sqrt{\frac{c}{2\pi}}\frac{e^{-\frac{c}{2x}}}{x^{\frac{3}{2}}}$ with $c>0$ for $x\in (0,\infty)$. It is straightforward to show that these conditions hold for L\'evy distributions with any  $c>0$.}, etc. Thus, when $\bar{v}$ is large,  the two-bidder robust regularity conditions hold for a distribution that is an approximation of some heavy-tailed distribution.

\subsection{$N$ Bidders}\label{s52}
The structure of this section is similar to Section \ref{s52}: I first define a mechanism  and a joint distribution. Then I define  regularity conditions under which the mechanism and the joint distribution form a saddle point (Theorem \ref{t2}). Then I illustrate Theorem \ref{t2}.\\
\indent \textit{The second-price auction with $\bar{v}$-scaled $Beta (\frac{1}{N-1},1)$ distributed random reserve} is defined as follows. The auctioneer first draws a random reserve $r$ from the $\bar{v}-$scaled $Beta (\frac{1}{N-1},1)$ distribution. That is, the cumulative distribution function of the random reserve $r$ is $G(r)=(\frac{r}{\bar{v}})^{\frac{1}{N-1}}$ with support $[0,\bar{v}]$. Then the $N$ bidders bid simultaneously.  The highest bidder  wins the object if her bid is also higher than $r$, and she pays the maximum of $r$ and the second highest bid; a bidder whose bid is not the highest loses the auction and pays nothing. In case of ties, each bidder wins the object with an equal probability if the bid is higher than $r$, and the winner pays the bid. \\
\indent  Equivalently, it can be defined by $(q^*,t^*)$ as follows. I denote the highest valuation  in a valuation profile $v$ by $v(1)$, and the second highest valuation (if any) by $v(2)$.   If   $\#\{k:v_k=v(1)\}=1$, then  $q^{**}_i(v)=(\frac{v(1)}{\bar{v}})^{\frac{1}{N-1}}, q^{**}_j(v)=0$ and $t^{**}_i(v)=\frac{v(1)^{\frac{N}{N-1}}+(N-1)v(2)^{\frac{N}{N-1}}}{N\bar{v}^{\frac{1}{N-1}}}, t^{**}_j(v)=0$ for $i\in \{k:v_k=v(1)\} $ and $j\notin \{k:v_k=v(1)\}$; if $\#\{k:v_k=v(1)\}=K\ge 2$, then  $q^{**}_i(v)=\frac{1}{K}(\frac{v(1)}{\bar{v}})^{\frac{1}{N-1}}, q^{**}_j(v)=0$ and $t^{**}_i(v)=\frac{v(1)^{\frac{N}{N-1}}}{K\bar{v}^{\frac{1}{N-1}}}, t^{**}_j(v)=0$ for $i\in \{k:v_k=v(1)\} $ and $j\notin \{k:v_k=v(1)\}$.\\
\indent The joint distribution $\pi^{**}$ is symmetric and is defined as follows\footnote{Here  $\pi^{**}(v)$ denote the density of the valuation profile $v$ whenever the density exists and $Pr^{**}(v)$ denote the probability mass of the valuation profile $v$ whenever there is some probability mass on $v$. The marginal distributions that  this result covers have a probability mass on the maximum valuation $\bar{v}$.  In the joint distribution $\pi^{**}$, there is a (non-negative) probability mass on the point $(\underbrace{\bar{v},\cdots, \bar{v}}_{N})$.}.
 The support of $\pi^{**}$ is  $V^+:=\{v\in V|\text{$ v_i=v(1)$ for any $i$ or $\exists i$ s.t. $v_i=v(1)>v_j=v(2)$ for any $j \neq i$}\}$. That is, $v\in V^+$ if either all bidders have the same valuations or there is a unique highest bidder and all of the remaining bidders have the same valuations.  If $v\notin V^+$, then $\pi^{**}(v)=0$.   If $v\in V^+$, then \[
\pi^{**}(v_i,v_{-i})= \left\{
\begin{array}{lll}
f(0)      &      & \\
\qquad \qquad\qquad \qquad \qquad\qquad\text{if $v=(0,
\cdots,0)$;}     &      & \\
0      &      & \\
\qquad \qquad\qquad \qquad \qquad\qquad\text{if $0=v_j<v_i,\forall j\neq i$;}     &      & \\
\frac{1}{(N-1)v(1)^2}(v(2)f(v(2))-\frac{v(2)^{-\frac{N}{N-1}}}{N-1}\int_0^{v(2)}x^{\frac{N}{N-1}}f(x)dx)       &      & \\
\qquad \qquad\qquad \qquad \qquad\qquad\text{if $0<v(2)=v_j\le v_i=v(1) < \bar{v},\forall j\neq i$;}     &      & \\
\frac{1}{(N-1)\bar{v}}(v(2)f(v(2))-\frac{v(2)^{-\frac{N}{N-1}}}{N-1}\int_0^{v(2)}x^{\frac{N}{N-1}}f(x)dx)     &      & \\
\qquad \qquad\qquad \qquad \qquad\qquad\text{if $0<v(2)=v_j<v_i= \bar{v},  \forall j\neq i$.}     &      & 
\end{array} \right. \]
\[Pr^{**}(\underbrace{\bar{v},\cdots, \bar{v}}_{N})=Pr(\bar{v})-\frac{\int_{(0,\bar{v})}x^{\frac{N}{N-1}}f(x)dx}{(N-1)\bar{v}^\frac{N}{N-1}}. \]
\indent \textit{The N-bidder robust regularity conditions (I)} are defined as follows: $f(x)\ge \frac{x^{-\frac{2N-1}{N-1}}}{N-1}\int_0^xs^{\frac{N}{N-1}}f(s)ds$  for $x\in(0,\bar{v})$ and $Pr(\bar{v}) \ge \frac{\int_{(0,\bar{v})}x^{\frac{N}{N-1}}f(x)dx}{(N-1)\bar{v}^\frac{N}{N-1}}$\footnote{With slight abuse of notation, here the condition for $Pr(\bar{v})$ is also referred to as the probability mass condition.}. The  \textit{N-bidder robust regularity conditions (II)} are defined as follows: $x^2f(x)$ is non-decreasing for $x\in(0,\bar{v})$ and $Pr(\bar{v}) \ge \frac{\int_{(0,\bar{v})}x^{\frac{N}{N-1}}f(x)dx}{(N-1)\bar{v}^\frac{N}{N-1}}$.
\begin{remark}\label{r3}
\normalfont Here the probability mass condition will  vanish as the number of the bidders goes to infinity. To see this,  note that 
$\frac{\int_{(0,\bar{v})}x^{\frac{N}{N-1}}f(x)dx}{(N-1)\bar{v}^\frac{N}{N-1}}\le \frac{\int_{(0,\bar{v})}\bar{v}^{\frac{N}{N-1}}f(x)dx}{(N-1)\bar{v}^\frac{N}{N-1}}\le  \frac{1}{N-1}\rightarrow 0 $ as $N\rightarrow \infty$. Therefore,  the probability mass condition is non-restrictive when the number of bidders is large.
\end{remark}
\begin{theorem}\label{t2}
For the N-bidder case ($N\ge 3$), the second-price auction with the $\bar{v}-$scaled $Beta (\frac{1}{N-1},1)$ distributed random reserve is a maxmin mechanism across standard dominant-strategy mechanisms if the $N-$bidder robust regularity conditions (I) hold. The revenue guarantee is $\frac{E[X^{\frac{N}{N-1}}]}{\bar{v}^{\frac{1}{N-1}}}$. The joint distribution $\pi^{**}$ is a worst-case correlation structure.  In addition,  the  $N-$bidder robust regularity conditions (II) imply the $N-$bidder robust regularity conditions (I).
\end{theorem}
\begin{remark}
\normalfont It is straightforward that the second-price auction with the $\bar{v}-$scaled $Beta (\frac{1}{N-1},1)$ distributed random reserve converges to the second-price auction without a reserve as the number of bidders goes to infinity. The asymptotic behaviour of the random reserve is   consistent with the empirical finding   that reserve
prices are substantially lower than the optimal ones  under the estimated
distribution of values (e.g., \cite{paarsch1997deriving}, \cite{mcafee2002set}, \cite{haile2003inference} and \cite{bajari2003winner}).
\begin{definition}
\normalfont Given a marginal distribution $F$, suppose the revenue guarantee of a maxmin mechanism across all dominant-strategy mechanisms is $Opt_N(F)$ for each  $N-$bidder case. Consider a  dominant-strategy mechanism  $M_N$ for each $N-$bidder case. Suppose the revenue guarantee of $M_N$ is $Reg_N(F)$.    I say $M_N$ is \textit{asymptotically optimal} across all dominant-strategy mechanisms given the marginal distribution $F$ if $Opt_N(F) - Reg_N(F)\to 0$ as $N\to \infty$. 
\end{definition}
\end{remark}
\begin{remark}\label{r4}
\normalfont The second-price auction with the  $\bar{v}-$scaled $Beta (\frac{1}{N-1},1)$ distributed random reserve is asymptotically optimal across all dominant-strategy mechanisms, regardless of the marginal distribution. Furthermore, the rate of convergence is $O(\frac{1}{N})$\footnote{In addition, this rate of convergence is the fastest across all standard dominant-strategy mechanisms, as is shown in \cite{he2020correlation}.}.  To see these, recall  that $E[X]$ is an upper bound of the revenue guarantee for any dominant-strategy mechanism. This is because  it is always possible that adversarial nature  chooses the maximally positively correlated distribution. But by the Dominated Convergence Theorem, I have that  $$
\frac{E[X^{\frac{N}{N-1}}]}{\bar{v}^{\frac{1}{N-1}}}\rightarrow E[X]$$as $N\rightarrow \infty$. Furthermore, let $j(x):=x-\frac{x^{\frac{N}{N-1}}}{\bar{v}^{\frac{1}{N-1}}}$. Because $j'(x)=1-\frac{Nx^{\frac{1}{N-1}}}{(N-1)\bar{v}^{\frac{1}{N-1}}}$ and $j''(x)=-\frac{Nx^{\frac{2-N}{N-1}}}{(N-1)^2\bar{v}^{\frac{1}{N-1}}}\le 0$, $j(x)$ is maximized at $x=(\frac{N-1}{N})^{N-1}\bar{v}$ and the maximized value is $(\frac{N-1}{N})^{N-1}\cdot \frac{\bar{v}}{N}$ by simple calculation. Then I have that $E[X]-\frac{E[X^{\frac{N}{N-1}}]}{\bar{v}^{\frac{1}{N-1}}}\le (\frac{N-1}{N})^{N-1}\cdot\frac{\bar{v}}{N}\le \frac{\bar{v}}{N}$. Therefore the rate of convergence is $O(\frac{1}{N})$. 
\end{remark}
Now I illustrate Theorem \ref{t2}. I start with the illustration of the mechanism. \\
\indent This mechanism exhibits a robust property in the following sense: for any conceivable joint distribution whose support is $V^+$, the expected revenue under this mechanism is the same. To see this, note that the total revenue from a valuation $v\in V^+$ is $t^{**}(v)=\sum_{i=1}^Nt^{**}_i(v)=\sum_{i=1}^N\frac{v_i^{\frac{N}{N-1}}}{N\bar{v}^{\frac{1}{N-1}}}$. Therefore, the expected revenue is $\frac{E[X^{\frac{N}{N-1}}]}{\bar{v}^{\frac{1}{N-1}}}$ for any conceivable joint distribution whose support is $V^+$. I will use the linear programming duality theorem to show that such an joint distribution indeed minimizes the expected revenue across all conceivable joint distributions. 
\begin{proposition}\label{p4}
For the N-bidder case ($N\ge 3$), under the second-price auction with the  $\bar{v}-$scaled $Beta (\frac{1}{N-1},1)$ distributed random reserve,  any conceivable joint distribution whose support is $V^+$ minimizes the expected revenue across all conceivable joint distributions. The minimized expected revenue is $\frac{E[X^{\frac{N}{N-1}}]}{\bar{v}^{\frac{1}{N-1}}}$.
\end{proposition}
\begin{proof}
The proof is in  Appendix \ref{ab} which presents the details about the construction of the mechanism.
\end{proof}
Proposition \ref{p4} implies that the constructed joint distribution $\pi^{**}$ minimizes the expected revenue across all conceivable joint distributions under the mechanism, as the support of the constructed joint distribution $\pi^{**}$ is $V^+$. \\
\indent The joint distribution $\pi^{**}$ is then obtained by a condition requiring that the highest bidder's virtual value be 0 except when her valuation is $\bar{v}$. Formally, \[
\phi_i^{**}(v_i,v_{-i})=0\quad \text{if} \quad v\in V^+,  \max_{j\neq i}v_j\le v_i<\bar{v}. \tag{1'}\label{1'}\]
\indent If the property \eqref{1'} holds for a joint distribution, then it is straightforward that the second-price auction with the $\bar{v}-$scaled $Beta (\frac{1}{N-1},1)$ distributed random reserve maximizes the expected revenue across all standard dominant-strategy mechanisms given this joint distribution. Note that I do not impose any condition on lower bidders' virtual values. This is because I  restrict attention to standard dominant-strategy mechanisms, and then  a bidder whose bid is not the highest is not allocated the object and pays nothing.  
\indent The $N-$bidder robust regularity conditions (I)   guarantee that the property \eqref{1'} holds  for the constructed joint distribution $\pi^{**}$. The $N-$bidder robust regularity conditions (II) are simpler conditions, and, as I will show,  they imply the $N-$bidder robust regularity conditions (I). 
\begin{proposition}\label{p5}
For the N-bidder case ($N\ge 3$), if the $N-$bidder robust regularity conditions (I) hold, then the second-price auction with the $\bar{v}$-scaled $Beta (\frac{1}{N-1},1)$ distributed random reserve maximizes the expected revenue across all standard dominant-strategy mechanisms under the joint distribution $\pi^{**}$. In addition, the $N-$bidder robust regularity conditions (II) imply  the $N-$bidder robust regularity conditions (I). 
\end{proposition}
\begin{proof}
The proof is in  Appendix \ref{ab} which presents details about the construction of the joint distribution $\pi^{**}$.
\end{proof}
\indent Theorem \ref{t2} follows immediately from Proposition \ref{p4} and Proposition \ref{p5}. \\

\indent Now I present the result for the special case: the equal-revenue distribution. In contrast to the two-bidder case, bidders' valuations are not independently distributed in the worst-case correlation structure for the $N-$bidders  case ($N\ge 3$).
\begin{corollary}\label{c4}
For the $N-$bidders case ($N\ge 3$),  if the marginal distribution is an  equal-revenue distribution $(\alpha \in (0,\bar{v}))$:  \[
F(v)=
\left\{
\begin{array}{rcl}
1-\frac{\alpha}{v}     &      & {\normalfont \text{if}\quad\alpha \le v < \bar{v};}\\
1      &      & {\normalfont \text{if}\quad v=\bar{v},}
\end{array} \right.\] 
then   the second-price auction with the $\bar{v}-$scaled $Beta (\frac{1}{N-1},1)$ distributed random reserve is a  maxmin mechanism across standard dominant-strategy  mechanisms. The revenue guarantee is $N\alpha-\frac{(N-1)\alpha^{\frac{N}{N-1}}}{\bar{v}^{\frac{1}{N-1}}}$.  The worst-case correlation structure is symmetric and is defined as follows. \[\pi(v_i,v_{-i})= \left\{
\begin{array}{lll}
\frac{(\frac{v(2)}{\alpha})^{-\frac{N}{N-1}}}{(N-1)v(1)^2}       &      & \\
\qquad \qquad\qquad \qquad \qquad\qquad \text{\normalfont if $\alpha \le v(2)=v_j \le v_i= v(1)<\bar{v},  \forall j\neq i$;}     &      & \\
\frac{(\frac{v(2)}{\alpha})^{-\frac{N}{N-1}}}{(N-1)\bar{v}}     &      & \\
\qquad \qquad\qquad \qquad \qquad\qquad\text{\normalfont if $\alpha \le v_j=v(2)<v_i=\bar{v},  \forall j\neq i$;}     &      & \\
0      &      &  \\
\qquad \qquad\qquad \qquad \qquad\qquad\text{\normalfont if otherwise.}     &      &
\end{array} \right. \]
\[Pr(\underbrace{\bar{v},\cdots, \bar{v}}_{N})=(\frac{\alpha}{\bar{v}})^{\frac{N}{N-1}}. \]
\end{corollary}
\begin{proof}
It is straightforward to show that an equal-revenue distribution satisfies the $N$-bidder robust regularity conditions (I) (and (II)). Then Theorem 
\ref{t2} implies Corollary \ref{c4}.
\end{proof}
\subsection{When Probability Mass Condition Fails}\label{ae}
When the probability mass condition fails, I characterize a maxmin mechanism across standard dominant-strategy mechanisms. 
\begin{theorem}\label{t5}
For the N-bidder case ($N\ge 2$), if $x^2f(x)$ is non-decreasing for $x\in (0,\infty)$ and $Pr(\bar{v})<\frac{\int_{(0,\bar{v})}x^{\frac{N}{N-1}}f(x)dx}{(N-1)\bar{v}^\frac{N}{N-1}}$, then the second-price auction with the $s^*-$scaled $Beta(\frac{1}{N-1},1)$ distributed random reserve is a maxmin mechanism across standard dominant-strategy mechanisms where $s^*\in (0,\bar{v})$ is a solution to \[
\frac{\int_{(0,s)}x^{\frac{N}{N-1}}dF(x)}{s^{\frac{N}{N-1}}}=(N-1)(1-F(s)).\tag{SD}\label{sd}\]
In addition, the revenue guarantee is $Ns^*(1-F(s^*))$.
\end{theorem}
\begin{proof}
The proof is in  Appendix \ref{ab}.
\end{proof}
\begin{remark}
\normalfont This result  generalizes and strengthens Theorem 3 in \cite{he2022correlation}. First, the conditions in this result  allow for a probability mass on the maximum valuation. Second, the condition for $x<\bar{v}$ is weaker than that $xf(x)$ is non-decreasing\footnote{It is straightforward that $xf(x)$ is non-decreasing implies that $x^2f(x)$ is non-decreasing, but not vice versa.}, required in their result.
\end{remark}
\indent Under this mechanism, the cumulative distribution function of the random reserve $r$  is $G_{s*}(r)=(\frac{r}{s^*})^{\frac{1}{N-1}}$ with support $[0,s^*]$. Therefore, the highest bidder will be allocated the object with a probability less than one if her valuation is less than $s^*$, and will be allocated the object with probability one   if her valuation is weakly higher than $s^*$. I follow the saddle point approach to show Theorem \ref{t5}. Notably, the constructed worst-case correlation structure exhibits the property that the highest bidder's virtual value is 0 when her valuation is less than $s^*$, and is weakly positive when her valuation is weakly higher than $s^*$. I relegate the details to  Appendix \ref{ab}.
\section{Robust Dominance}\label{s6}
I have shown that the second-price auction with the $\bar{v}-$scaled $Beta (\frac{1}{N-1},1)$ distributed random reserve is a maxmin mechanism   under certain regularity conditions. Then what if the regularity conditions do not hold? How does this mechanism perform? As a first topic of this section,  I compare the performance of this mechanism with that of the posted-price mechanism, which is a dominant-strategy mechanism commonly  used in practice. For exposition, I assume that  the marginal distribution admits a positive density function everywhere on $[0,\bar{v}]$  in this section. I denote the set of all such distributions by $\Delta^c[0,\bar{v}]$.
\begin{definition}
 I say a mechanism $M$  \textit{dominates}  a family of mechanisms $\mathcal{M}$ for a set  of marginal distributions if for any marginal distribution in this set,  the revenue guarantee of $M$ is strictly greater than that of any mechanism in the family $\mathcal{M}$ .
\end{definition}
\begin{definition}
\normalfont  I say 
a family of mechanisms $\mathcal{M}_1$ \textit{universally dominates} another family of mechanisms $\mathcal{M}_2$ if for any  $F\in \Delta^c[0,\bar{v}]$, there exists a mechanism in $\mathcal{M}_1$ with a revenue guarantee strictly greater than that of any mechanism in $\mathcal{M}_2$.
\end{definition}
\begin{proposition}\label{p6}
For any $N\ge 2$, the second-price auction with the $\bar{v}-$scaled $Beta (\frac{1}{N-1},1)$ distributed random reserve  dominates the family of  posted-price mechanisms for the set $\mathcal{H}=\{F\in \Delta^c[0,\bar{v}]|\text{the revenue function $x\cdot(1-F(x))$ is strictly concave}\}$.
\end{proposition}
\begin{proof}
The proof is in  Appendix \ref{ac}. 
\end{proof}
\indent Motivated by  the main idea embedded in the construction, I propose a family of second-price auctions with $t-$scaled $Beta (\frac{1}{N-1},1)$ distributed random reserves where $t\in (0,\bar{v})$, denoted by $\mathcal{M}_{SP-\beta}$. As a second topic of this section, I study the performance of the proposed family of auctions.  Formally, the cumulative distribution function of the random reserve $r$ in this family is $
G_t(r)=(\frac{r}{t})^{\frac{1}{N-1}}$  with support $[0, t]$ for some $t\in (0,\bar{v})$. Under such a random reserve, if the valuation of the highest bidder is above the  threshold $t$, the object will be fully allocated to her. For each $t$, I am able to identify a non-trivial lower bound of the revenue guarantee by constructing a set of feasible dual variables.
\begin{lemma}\label{l1}
A lower bound of the revenue guarantee of the second-price auction with $t-$scaled $Beta (\frac{1}{N-1},1)$ distributed random reserve where $t\in (0,\bar{v})$  is  \[\int_{(0,t)}\frac{x^{\frac{N}{N-1}}}{t^{\frac{1}{N-1}}}dF(x)+t(1-F(t)). \]
\end{lemma}
\begin{proof}
The proof is in  Appendix \ref{ac}.
\end{proof}
\indent  Lemma \ref{l1} suggests a potential criterion under which the auctioneer selects an auction from this family. Although the revenue guarantee of an auction in this family may depend on the details of the marginal distribution and thus may be hard to be identified, it has a non-trivial lower bound. Then I can  compare this lower bound  with the revenue guarantees of  some  dominant-strategy mechanisms commonly used in practice. 
\begin{theorem}\label{t3}
For any $N\ge 2$, $\mathcal{M}_{SP-\beta}$  universally dominates the family of posted-price mechanisms. 
\end{theorem}
\begin{proof}
The proof is in  Appendix \ref{ac}.
\end{proof}
\begin{theorem}\label{t4}
For any $N\ge 2$, $\mathcal{M}_{SP-\beta}$  universally dominates the family of second-price auctions with non-negative deterministic reserves. 
\end{theorem}
\begin{proof}
The proof is in  Appendix \ref{ac}.
\end{proof}
\indent Strikingly,  I do not require any distributional assumption for these two theorems to hold. Moreover,  Theorem \ref{t3} and \ref{t4} imply that for a given marginal distribution, the auctioneer can find  a  second-price auction with a $t-$scaled $Beta (\frac{1}{N-1},1)$ distributed random reserve whose revenue guarantee is strictly higher than that of any posted-price mechanism and  any second-price auction with a non-negative deterministic reserve.  In addition,  Theorem \ref{t3} can be interpreted as that the competition effect dominates the  adversarial correlation effect. To see this, note that Theorem \ref{t3} implies that there exists an  auction for the two-bidder case  that generates strictly higher  revenue guarantee than the monopoly revenue from one bidder, regardless of the marginal distribution. Thus, even if  nature picks the worst-case correlation structure, it is always \textit{strictly} more desirable for the auctioneer to have just one more bidder.
\section{Concluding Remarks}\label{s7}
In this paper, I consider the correlation-robust framework and  show, among others, that the second-price auction with the uniformly distributed random reserve maximizes the revenue guarantee across all dominant-strategy mechanisms for the two-bidder case, and  that the second-price auction with the $Beta (\frac{1}{N-1},1)$ distributed random reserve maximizes the revenue guarantee across standard dominant-strategy mechanisms for the $N$-bidder ($N\ge 3$) case. These auctions have familiar formats and admit simple descriptions which  do not require the information of the marginal distribution except for the maximum valuation of of a generic bidder.  Thus, these auctions are more practical compared  with, for example,  Myerson's auction, which often requires the full information of the marginal distribution to calculate the optimal reserve.
To my knowledge, this paper is the first to characterize a  maxmin mechanism across \textit{all} dominant-strategy mechanism in the correlation-robust framework. It remains an  open question what the maxmin mechanisms across all dominant-strategy  mechanisms are for general number of bidders. The constructive method may shed light on other robust design problems and general robust optimization problems. 

\bibliographystyle{apalike}
\bibliography{abc}

\appendix
\section{Proof for Section \ref{s4}: Proposition \ref{p1}}\label{aa}
1)  Dominant-strategy incentive compatibility (DSIC) for a type $v_i$ requires that for any $v_i'\neq v_i$: $$v_iq_i(v_i,v_{-i}) - t_i(v_i,v_{-i})\ge v_iq_i(v_i',v_{-i}) - t_i(v_i',v_{-i}). $$
DSIC also requires that:$$ v_i'q_i(v_i',v_{-i}) - t_i(v_i',v_{-i})\ge v_i'q_i(v_i,v_{-i}) - t_i(v_i,v_{-i}).$$Adding the two inequalities, I obtain that $$(v_i-v_i')(q_i(v_i,v_{-i})-q_i(v_i',v_{-i}))\ge 0.$$It follows that $q_i(v_i,v_{-i})\ge q_i(v_i',v_{-i})$ whenever $v_i>v_i'$ .\\
2) Fix any $v_{-i}$, and define $$U_i(v_i)=v_iq_i(v_i,v_{-i})-t_i(v_i,v_{-i}).$$ 
By the two inequalities in 1), I obtain that $$(v_i'-v_i)q_i(v_i,v_{-i})\le U_i(v_i')-U_i(v_i)\le (v_i'-v_i)q_i(v_i',v_{-i}).$$ Dividing throughout by $v_i'-v_i$, I obtain that $$q_i(v_i,v_{-i})\le \frac{U_i(v_i')-U_i(v_i)}{(v_i'-v_i)}\le q_i(v_i',v_{-i}).$$As $v\uparrow v'$,  I have that $$
\frac{dU_i(v_i)}{dv_i}=q_i(v_i,v_{-i}).$$Then I obtain that $$t_i(v_i,v_{-i})= v_iq_i(v_i,v_{-i})-\int_{0}^{v_i}q_i(s,v_{-i})ds-U_i(0).$$
Note that $U_i(0)\ge 0$ by the ex-post individually rational constraint. If $U_i(0)>0$, then I can reduce it to 0 so that I can increase the revenue from any valuation profile in which the others' valuation profile is $v_{-i}$.  And the revenue guarantee will be weakly greater. Thus, when searching for a maxmin mechanism, it is without loss to let $U_i(0)=0$. Then I obtain that $t_i(v_i,v_{-i})=v_iq_i(v_i,v_{-i})-\int_{0}^{v_i}q_i(s,v_{-i})$.
\section{ Proofs for Section \ref{s5}}\label{ab}
\subsection{Proof of Proposition \ref{p3}}\label{ab1}
First, I illustrate the details about the construction of $\pi^*$. Note that by allocating all marginal density $f(0)$  to valuation profile $(0,0)$, I have that  $\phi^*_i(v_i,0)=\phi^*_j(0,v_j)=0$ for any $v_i$ and $v_j$. Thus, the property \eqref{1} trivially holds for any one of these valuation profiles. Now
let $A_{kj}:=\{v|k\le v_1\le j,v_2=k\}$, and define $c(0):= f(0)$ and  $c(k):=\int_{A_{k\bar{v}}}d\pi^*$ for $k>0$. Consider the valuation profile $(v_1,v_2)$ where $0<v_2\le v_1< \bar{v}$. In order for the virtual value to satisfy the property \eqref{1}, I  have that \[
\phi_1^*(v_1,v_2)=v_1-\frac{c(v_2)-\int_{v_2}^{v_1}\pi^*(x,v_2)dx}{\pi^*(v_1,v_2)}=0,\quad \forall 0<v_2\le v_1<\bar{v}.\]These equations are essentially a system of  ordinary differential equations, whose solution is well known\footnote{The solution is reminiscent of the equal-revenue distribution.}:
\[\label{eq12}
   \pi^*(v_1,v_2)=\frac{v_2c(v_2)}{v_1^2}, \quad \forall 0<v_2 \le v_1 <\bar{v}, \tag{B.1}
\]
\[\label{eq13}
  \pi^*(\bar{v},v_2)=\frac{v_2c(v_2)}{\bar{v}}, \quad \forall 0<v_2  < \bar{v}. \tag{B.2} 
\]
By symmetry, I also obtain $\pi^*(v_2,v_1)=\pi^*(v_1,v_2)$ for $0<v_2 \le v_1 <\bar{v}$  and $\pi^*(v_2,\bar{v})=\pi^*(\bar{v},v_2)$ for $0<v_2  < \bar{v}$ . Finally,
\[\label{eq14}
   Pr^*(\bar{v},\bar{v})=Pr(\bar{v})-\frac{\int_{j\in(0,\bar{v})}jc(j)dj}{\bar{v}}. \tag{B.3}
\]
Now I solve for $c(k)$. Note since the marginal distribution is the same across the two bidders, given the above derivation, $c(k)$ must satisfy the following condition:
\[\label{eq15}
   c(k) = f(k)-\frac{\int_{0}^{k}jc(j)dj}{k^2}, \quad \forall 0< k < \bar{v}. \tag{B.4}
\]
To see this, note $f(k)=\int_{\{0\le v_1\le \bar{v}, v_2=k\}}d\pi^*=\int_{A_{k\bar{v}}\cup \{0\le v_1<k,v_2=k\}}d\pi^*=\int_{A_{k\bar{v}}}d\pi^*+\int_{\{0\le v_1<k,v_2=k\}}d\pi^*=\int_{A_{k\bar{v}}}d\pi^*+\int_{\{v_1=k, 0\le v_2<k\}}d\pi^*$ where the last equality follows from symmetry. Multiplying  both sides of \eqref{eq15} by $k$, I obtain that \[kc(k)=kf(k)-\frac{\int_{0}^{k}jc(j)dj}{k},\quad \forall 0< k < \bar{v}. \]
Define $g(k):=\int_{0}^{k}jc(j)dj$ for $0<k<\bar{v}$. Then I have that  \[g'(k)=kf(k)-\frac{g(k)}{k}, \quad \forall 0< k < \bar{v}. \]
Note this is an ordinary differential equation, and  I solve for $g(k)$:
\[\label{eq16}
   g(k)=\frac{1}{k}\int_0^kj^2f(j)dj, \quad \forall 0< k < \bar{v}. \tag{B.5}
\]
From this I compute $c(k)$: 
\[\label{eq17}
c(k)=f(k)-\frac{\int_{0}^{k}j^2f(j)dj}{k^3},   \quad \forall 0< k < \bar{v}. \tag{B.6}
\]
Plugging \eqref{eq17} to \eqref{eq12}, \eqref{eq13} and \eqref{eq14}, I obtain the joint distribution $\pi^*$.\\

\indent To guarantee that it is possible to construct $\pi^*$, it has to be a feasible joint distribution in that the density (or probability mass) has to be non-negative for all valuation profiles, i.e., $\pi^*(v_1,v_2)\ge 0$ for $0\le v_1,v_2 <\bar{v}$ and $Pr^*(\bar{v},\bar{v})\ge 0$. Therefore, I have that 
\[\label{eq18}
 f(k)-\frac{\int_{0}^{k}j^2f(j)dj}{k^3}\ge 0, \quad \forall 0<k<\bar{v},   \tag{B.7}
\]
\[\label{eq19}
   Pr(\bar{v})\ge \frac{\int_{x\in(0,1)}x^2f(x)dx}{\bar{v}^2}. \tag{B.8}
\]
Now I show that the first part of the two-bidder robust regularity conditions implies \eqref{eq18}. To see this, note that if $x^2f(x)$ is non-decreasing  for $x\in (0,\bar{v})$, then for any $0<k<\bar{v}$, I have that  \[
f(k)-\frac{\int_{0}^{k}j^2f(j)dj}{k^3}\ge f(k)-\frac{\int_{0}^{k}k^2f(k)dj}{k^3}= 0,\label{eq20}\tag{B.9}\]
where the inequality follows from that $j^2f(j)\le k^2f(k)$ if $j\le k$.\\
\indent Now given that the construction is feasible,  I argue that the two-bidder robust regularity conditions guarantee that the property \eqref{2} holds. Given that the property \eqref{1} holds for $\pi^*$, it suffices to show $$\phi_2(v_1, v_2)\le 0$$ if $0< v_2\le v_1< \bar{v}$. I now calculate $\phi_2(v_1,v_2)$ for $0< v_2\le v_1< \bar{v}$:\[
\begin{split}
 \phi_2(v_1, v_2)&=v_2-\frac{f(v_1)-\int_0^{v_2}\pi^*(v_1,t)dt}{\pi^*(v_1,v_2)}\\ &= v_2-\frac{f(v_1)-\int_0^{v_2}c(t)\frac{t}{v_1^2}dt}{c(v_2)\frac{v_2}{v_1^2}}\\ &=v_2-\frac{f(v_1)-\frac{1}{v_1^2v_2}\int_0^{v_2}t^2f(t)dt}{(f(v_2)-\frac{\int_{0}^{v_2}s^2f(s)ds}{v_2^3})\frac{v_2}{v_1^2}}, 
\end{split}\]  
where the second equality follows from \eqref{eq12} and the third equality follows from \eqref{eq16} and \eqref{eq17}.
Now it is straightforward  that for any $0< v_2\le v_1< \bar{v}$: $$\phi_2(v_1, v_2)\le 0 \Longleftrightarrow 
v_2^2f(v_2)\le v_1^2f(v_1).$$

\subsection{Proof of Proposition  \ref{p4}}
First, I illustrate the details about the construction of the mechanism. I first write down the primal minimization problem for adversarial nature given a mechanism $(q,t)$ and derive its dual maximization problem.  Formally, let   $\{\lambda_i(v_i)\}_{i\in\{1,2,\cdots, N\},v_i\in[0,\bar{v}]} $ be dual variables. \[(P)\quad\inf_{\pi\in\Pi(F)}\int_{v\in [0,\bar{v}]^N}\sum_{i=1}^N t_i(v)d\pi(v)\] subject to \[
\int_{[0,\bar{v}]^{N-1}}d\pi(v_i,v_{-i})=f(v_i),\quad \forall  v_i\in[0,\bar{v}),
\] \[
\int_{[0,\bar{v}]^{N-1}}d\pi(v_i=\bar{v},v_{-i})=Pr(v_i=\bar{v}).
\] 
\[
(D)\quad \sup_{\{\lambda_i(v_i)\}} \sum_{i=1}^N\int_{0}^{\bar{v}}\lambda_i(v_i)dF(v_i)\]
subject to
\[\label{eq27}
  \sum_{i=1}^N \lambda_i(v_i)\le \sum_{i=1}^N t_i(v),\quad \forall v\in [0,\bar{v}]^N.  \tag{B.10}
\] 
It is straightforward to show that weak duality holds \footnote{See, for example, \cite{he2022correlation}.}.   
The mechanism is constructed by  a complementary slackness condition as follows.
\[\label{eq28}
 \sum_{i=1}^N \lambda_i(v_i)= \sum_{i=1}^N t_i(v),\quad \forall v\in V^+.  \tag{B.11}
\]
I assume that $\lambda_i=\lambda$ for all $i\in I$, and  that the mechanism is a second-price auction with a random reserve  whose cumulative distribution function is $G$, then  \eqref{eq28} implies
\[\label{eq29}
    N\lambda(v_i)=v_iG(v_i),\quad \forall v_i\in [0,\bar{v}],\tag{B.12}
\]
\[\label{eq30}
    \lambda(v(1))+(N-1)\lambda(v(2))=v(1)G(v(1))-\int_{v(2)}^{v(1)}G(s)ds, \quad \forall  0\le v(2)<v(1)\le 1.\tag{B.13}
\]
Note by \eqref{eq29}, I have that for $v_i\in [0,\bar{v}]$,
\[\label{eq31}
    \lambda(v_i)=\frac{v_iG(v_i)}{N}.\tag{B.14}
\]
Plugging \eqref{eq31} to \eqref{eq30}, I obtain that for $0\le v(2)<v(1)\le \bar{v}$,
\[\label{eq32}
\frac{v(1)G(v(1))+(N-1)v(2)G(v(2))}{N}=v(1)G(v(1))-\int_{v(2)}^{v(1)}G(s)ds.  \tag{B.15}
\]
Taking first order derivatives with respect to $v(1)$ and $v(2)$, I obtain the same ordinary differential equation that for $x\in [0,\bar{v}] $,
\[\label{eq33}
    (N-1)xG'(x)=G(x). \tag{B.16}
\]
Given that $G$ is a distribution,  the solution to \eqref{eq33} is
\[
    G(x)=(\frac{x}{\bar{v}})^{\frac{1}{N-1}},\quad \forall x\in [0,\bar{v}].
\]
This is the $\bar{v}-$scaled $Beta(\frac{1}{N-1},1)$ distribution.\\
\indent Now I show that under the second-price auction with the $\bar{v}-$scaled $Beta(\frac{1}{N-1},1)$ distributed random reserve, any conceivable joint distribution with the support $V^+$ minimizes the expected revenue across all conceivable joint distributions. As I have argued in Section \ref{s52}, the expected revenue given such a joint distribution is $\frac{E[X^{\frac{N}{N-1}}]}{\bar{v}^{\frac{1}{N-1}}}$, then it suffices to show the value of $(D)$ is also $\frac{E[X^{\frac{N}{N-1}}]}{\bar{v}^{\frac{1}{N-1}}}$. To this end, I construct the dual variables as follows. For all $i\in\{1,2,\cdots,N\}$ and $x\in[0,1]$,
\[\label{eq34}
    \lambda_i(x)=\frac{x^{\frac{N}{N-1}}}{N\bar{v}^{\frac{1}{N-1}}}.
\]
Note that $\sum_{i=1}^N\int_{0}^{\bar{v}}\lambda_i(v_i)dF(v_i)=\frac{E[X^{\frac{N}{N-1}}]}{\bar{v}^{\frac{1}{N-1}}}$ under the constructed dual variables. Then, it suffices to show that  \eqref{eq27} holds  under the constructed dual variables. To see this, I divide valuation profiles into two cases.\\
\textit{Case 1}: $\#\{k:v_k=v(1)\}=1$.\\
In this case, $t(v)=\frac{v(1)^{\frac{N}{N-1}}+(N-1)v(2)^{\frac{N}{N-1}}}{N\bar{v}^{\frac{1}{N-1}}}$. The L.H.S. of \eqref{eq27} is maximized when all bidders except the highest bidder have the same valuations, and \eqref{eq27} holds with equality when the L.H.S. of \eqref{eq27} is maximized. Therefore \eqref{eq27} holds. \\
\textit{Case 2}: $\#\{k:v_k=v(1)\}\ge 2$.\\
In this case, $t(v)=\frac{v(1)^{\frac{N}{N-1}}}{\bar{v}^{\frac{1}{N-1}}}$. The L.H.S. of \eqref{eq27} is maximized when all bidders  have the same valuations, and \eqref{eq27} holds with equality when the L.H.S. of \eqref{eq27} is maximized.  Therefore \eqref{eq27} holds.
\subsection{Proof of Proposition \ref{p5}}
First, I illustrate the details about the construction of $\pi^{**}$. Note that by allocating all marginal density $f(0)$  to the valuation profile $\underbrace{(0,\cdots,0)}_{N}$, I have that  $\phi^{**}_i(v)=0$ for any $i$ and $v_i\ge 0, v_j=0,  \forall j\neq i$. Thus, the property \eqref{1'} trivially holds for any one of these valuation profiles. Now 
let $B_{kj}:=\{v|k\le v_1\le j,v_i=k,\quad\forall i\neq 1\}$, and define  $d(0):= f(0)$ and  $d(k):=\int_{B_{k\bar{v}}}d\pi^*$ for $k>0$. Consider the valuation profile $(v_1,\underbrace{v_2,\cdots,v_2}_{N-1})$ where $0<v_2\le v_1< \bar{v}$. In order for the virtual value of bidder 1 to satisfy the property \eqref{1'}, I have that $$
\phi^{**}_1(v_1,\underbrace{v_2,\cdots,v_2}_{N-1})=v_1-\frac{d(v_2)-\int_{v_2}^{v_1}\pi^{**}(s,\underbrace{v_2,\cdots,v_2}_{N-1})ds}{\pi^{**}(v_1,\underbrace{v_2,\cdots,v_2}_{N-1})}=0,\quad \forall 0<v_2\le v_1<\bar{v}.$$ These equations are essentially a system of  ordinary differential equations, whose solution is well known:
\[\label{eq21}
   \pi^{**}(v_1,\underbrace{v_2,\cdots,v_2}_{N-1})=\frac{v_2d(v_2)}{v_1^2}, \quad \forall 0<v_2 \le v_1 <\bar{v}, \tag{B.17}
\]
\[\label{eq22}
  \pi^{**}(1,\underbrace{v_2,\cdots,v_2}_{N-1})=\frac{v_2d(v_2)}{\bar{v}}, \forall \quad 0<v_2  < \bar{v}. \tag{B.18}
\]
By symmetry, I also obtain that $\pi^{**}(v)=\pi^{**}(v_1,\underbrace{v_2,\cdots,v_2}_{N-1})$ for $0<v_j=v_2 \le v_i=v_1 <\bar{v}, \forall j\neq i, \forall i$  and $\pi^{**}(v)=\pi^*(1,\underbrace{v_2,\cdots,v_2}_{N-1})$ for $0<v_j=v_{2}<v_i=\bar{v}, \forall j\neq i, \forall i$. Finally,
\[\label{eq23}
   Pr^{**}(\underbrace{\bar{v},\cdots, \bar{v}}_{N})=Pr(\bar{v})-\frac{\int_{j\in(0,\bar{v})}jd(j)dj}{\bar{v}}.\tag{B.19}
\]
Now I solve for $d(k)$. Note that $d(k)$ must satisfy the following condition:
\[\label{eq24}
   f(k)=(N-1)d(k)+\frac{\int_{0}^{k}jd(j)dj}{k^2}, \quad \forall 0< k < \bar{v}.
\]
To see this, suppose that the bidder 1's valuation is $k$. Then either $k$ is the highest valuation and other bidders all have a valuation of $j\in[0,k]$ (with a probability of $\frac{\int_{0}^{k}jd(j)dj}{k^2}$) or $k$ is the second highest valuation and one of the other bidders has the highest valuation (with a probability of $(N-1)d(k)$). Multiplying  both sides of \eqref{eq24} by  $k$, I obtain that \[kf(k)=(N-1)kd(k)+\frac{\int_{0}^{k}jd(j)dj}{k},\quad \forall 0< k <  \bar{v}. \]
Define $h(k):=\int_{0}^{k}jd(j)dj$ for $0<k< \bar{v}$. Then I have that  $$kf(k)=(N-1)h'(k)+\frac{h(k)}{k}, \quad \forall 0< k <  \bar{v}. $$
Note that this is an ordinary differential equation, and  I solve for $h(k)$:
\[\label{eq25}
   h(k)=\frac{\int_0^kj^{\frac{N}{N-1}}f(j)dj}{(N-1)k^{\frac{1}{N-1}}}, \quad \forall 0< k <  \bar{v}. \tag{B.20}
\]
From this I compute $d(k)$:
\[\label{eq26}
d(k)=\frac{1}{N-1}(f(k)-\frac{\int_{0}^{k}j^{\frac{N}{N-1}}f(j)dj}{(N-1)k^{1+\frac{N}{N-1}}}),    \quad \forall 0< k <  \bar{v}.  \tag{B.21}
\]
Plugging \eqref{eq26} to \eqref{eq21},\eqref{eq22} and \eqref{eq23}, I obtain the joint distribution $\pi^{**}$.\\
\indent To guarantee that $\pi^*$
is  a feasible joint distribution in  that the density (or probability mass) has to be non-negative for all valuation profiles, it is straightforward that the $N-$bidder robust regularity conditions (I) have to hold. Now I  show that the $N-$bidder robust regularity conditions (II) imply the $N-$bidder robust regularity conditions (I). To see this, note that if $x^2f(x)$ is non-decreasing  for $x\in (0,\bar{v})$, then for any $0<k<\bar{v}$, I have that  \[
f(k)-\frac{\int_{0}^{k}j^{\frac{N}{N-1}}f(j)dj}{(N-1)k^{1+\frac{N}{N-1}}}\ge f(k)-\frac{\int_{0}^{k}j^{\frac{N}{N-1}-2}k^2f(k)dj}{(N-1)k^{1+\frac{N}{N-1}}}= 0,\label{b22}\tag{B.22}\]
where the inequality follows from that $j^2f(j)\le k^2f(k)$ if $j\le k$.
\subsection{Proof of Theorem \ref{t5}}\label{ab4}
\begin{lemma}\label{l2}
If  $x^2f(x)$ is non-decreasing for $x\in (0,\bar{v})$ and $Pr(\bar{v})<\frac{\int_{(0,\bar{v})}x^{\frac{N}{N-1}}f(x)dx}{(N-1)\bar{v}^\frac{N}{N-1}}$, then there exists  $s^*\in (0,\bar{v})$ that  is a solution to \eqref{sd}.
\end{lemma}
\begin{proof}
 First,
note that if $s \uparrow \bar{v}$, the R.H.S of \eqref{sd} converges to $(N-1)Pr(\bar{v})$, thus the L.H.S. of \eqref{sd}> the R.H.S of \eqref{sd}.\\
\indent Next, take a monotone sequence $\{s_n\}_{n\in \mathbb{N}}$ where  $s_n\downarrow 0$ as $n\to \infty$, $s_1\in (0,\bar{v})$ and $\frac{s_{n+1}}{s_n}\le \frac{1}{2}$ for any $n$. \footnote{For example, $s_n=\frac{\bar{v}}{2^{n}}$ for $n\in \mathbb{N}$. } I will prove that $\limsup_{n\to \infty}s_nf(s_n)=0$ by contradiction.  Suppose that  $\limsup_{n\to \infty}s_nf(s_n)=c>0$, then  for any $ \epsilon>0$,  there exists a subsequence $\{s_{n_k}\}$ such that  $s_{n_k}f(s_{n_k})-c \ge \epsilon$ for any $k$. So $f(s_{n_k})\ge \frac{c-\epsilon}{s_{n_k}}$ for any $k$.
Let $\epsilon$ be $\frac{c}{2}$.  That $x^2f(x)$ is non-decreasing implies that for any $x\in (s_{n_{k+1}},s_{n_k})$, $f(x)\ge \frac{s_{n_{k+1}}^2f(s_{n_{k+1}})}{x^2}\ge \frac{s_{n_{k+1}}(c-\epsilon)}{x^2}=\frac{cs_{n_{k+1}}}{2x^2}$ for any $k$. Therefore $\int_{s_{n_{k+1}}}^{s_{n_{k}}}f(x)\ge \frac{c\cdot s_{n_{k+1}}}{2}(\frac{1}{s_{n_{k+1}}}-\frac{1}{s_{n_{k}}})\ge \frac{c}{4} $. Thus,  $\int_0^{\bar{v}}dF(x)\ge \sum_{k=1}^{K} \int_{s_{n_{k+1}}}^{s_{n_{k}}}f(x)\ge \frac{cK}{4}\to \infty$ as $K\to \infty$, a contradiction to the fact that $F$ is a probability measure. Therefore $\limsup_{n\to \infty}s_nf(s_n)=0$. This implies that $\lim_{n\to \infty}s_nf(s_n)=0$.  Now, by  L'H\^opital's rule,  \[
\begin{split}
    \lim_{n\to\infty}\frac{\int_{(0,s_n)}x^{\frac{N}{N-1}}dF(x)}{s_n^{\frac{N}{N-1}}}& = \lim_{n\to \infty}\frac{s_n^{\frac{N}{N-1}}f(s_n)}{\frac{N}{N-1}s_n^{\frac{1}{N-1}}}\\ &=\lim_{n\to \infty}\frac{(N-1)s_nf(s_n)}{N}\\ &=0.
\end{split}\]
Then, if $s_n\downarrow 0$, the L.H.S. of \eqref{sd} < the R.H.S of \eqref{sd}. By the Intermediate Value Theorem, there exists  $s^*\in (0,\bar{v})$ that  is a solution to \eqref{sd}.
\end{proof}
\begin{proposition}\label{p9}
If  $x^2f(x)$ is non-decreasing for $x\in (0,\bar{v})$ and $Pr(\bar{v})<\frac{\int_{(0,\bar{v})}x^{\frac{N}{N-1}}f(x)dx}{(N-1)\bar{v}^\frac{N}{N-1}}$, then the revenue guarantee of the second-price auction with the $s^*-$scaled $Beta(\frac{1}{N-1},1)$ distributed random reserve is at least $(N-1)s^*(1-F(s^*))$.
\end{proposition}
\begin{proof}
This follows immediately from  Lemma \ref{l1} and Lemma \ref{l2}.
\end{proof}
\begin{proposition}\label{p10}
If  $x^2f(x)$ is non-decreasing for $x\in (0,\bar{v})$ and $Pr(\bar{v})<\frac{\int_{(0,\bar{v})}x^{\frac{N}{N-1}}f(x)dx}{(N-1)\bar{v}^\frac{N}{N-1}}$, then there exists a joint distribution $\pi^{***}\in \pi(F)$ under which the second-price auction with the $s^*-$scaled $Beta(\frac{1}{N-1},1)$ distributed random reserve maximizes the expected revenue across standard dominant-strategy mechanisms. In addition, the maximized expected revenue is $(N-1)s^*(1-F(s^*))$.
\end{proposition}
\begin{proof}
The joint distribution $\pi^{***}$ is symmetric and is defined as follows. The support of $\pi^{***}$ is  $V^+$.   If $v\notin V^+$, then $\pi^{***}(v)=0$.   If $v\in V^+$, then \[
\pi^{***}(v_i,v_{-i})\footnote{The density function $\pi^{***}$ in the region $(0,\bar{v})^N$ is similar to the density function $\eta_F^*$ in the region $(0,1)^N$ in \cite{he2022correlation}. }= \left\{
\begin{array}{lll}
f(0)      &      & \\
\qquad \qquad\qquad \qquad \qquad\qquad\text{if $v=(0,
\cdots,0)$;}     &      & \\
0      &      & \\
\qquad \qquad\qquad \qquad \qquad\qquad\text{if $0=v_j<v_i,\forall j\neq i$;}      &      & \\
\frac{1}{(N-1)v(1)^2}(v(2)f(v(2))-\frac{v(2)^{-\frac{N}{N-1}}}{N-1}\int_0^{v(2)}x^{\frac{N}{N-1}}f(x)dx)       &      & \\
\qquad \qquad\qquad \qquad \qquad\qquad\text{if $0<v(2)=v_j\le v_i=v(1) \le s^*,\forall j\neq i$;}      &      & \\
\frac{f(v(1))}{(N-1)s^*(1-F(s^*))}(v(2)f(v(2))-\frac{v(2)^{-\frac{N}{N-1}}}{N-1}\int_0^{v(2)}x^{\frac{N}{N-1}}f(x)dx)       &      & \\
\qquad \qquad\qquad \qquad \qquad\qquad \text{if $0<v(2)=v_j \le s^*<v_i<\bar{v},\forall j\neq i$;}      &      & \\
\frac{Pr(\bar{v})}{(N-1)s^*(1-F(s^*))}(v(2)f(v(2))-\frac{v(2)^{-\frac{N}{N-1}}}{N-1}\int_0^{v(2)}x^{\frac{N}{N-1}}f(x)dx)       &      & \\
\qquad \qquad\qquad \qquad \qquad\qquad \text{if $0<v(2)=v_j\ \le s^*<v_i=\bar{v},\forall j\neq i$.}      &      & 
\end{array} \right. \]
It is straightforward to verify that $\pi^{***}\in \Pi(F)$. When $v_i=v(1)\le s^*$, the density function coincides with $\pi^{**}$. Therefore by the proof of Proposition \ref{p5}, \[ \phi_i^{***}(v)=0 \quad \text{for $v\in V^+, v_i=v(1)\le s^*$.}\] Note that under $\pi^{***}$,  when $v_i=v(1)>s^*$, $v_i$ and $v_{-i}$ are independent. Therefore $\phi_i^{***}(v)=v_i-\frac{1-F(v_i)}{f(v_i)}$ for $s^*<v_i=v(1)<\bar{v}$ and $\phi_i^{***}(v)=\bar{v}$ for $v_i=v(1)=\bar{v}$.\\
\indent Now I show that $\phi_i^{***}(v)\ge 0$ for $s^*<v_i=v(1)<\bar{v}$. First I show that $1-F(x)-xf(x)$ is non-increasing if $x^2f(x)$ is non-decreasing. To see this, note that for any $0<x_1\le x_2<\bar{v}$, \[
\begin{split}
    1-F(x_2)-x_2f(x_2)-[1-F(x_1)-x_1f(x_1)]&=x_1f(x_1)-x_2f(x_2)-\int_{x_1}^{x_2}f(x)dx\\ &\le x_1f(x_1)-x_2f(x_2)-\int_{x_1}^{x_2}\frac{x_1^2f(x_1)}{x^2}dx\\ &=\frac{x_1^2f(x_1)}{x_2}-x_2f(x_2)\\
    & \le 0,
\end{split}\]
where the first inequality follows from that  $x^2f(x)\ge x_1^2f(x_1)$ for $x_1\le x \le x_2$ and the second inequality follows from that  $x_2^2f(x_2)\ge x_1^2f(x_1)$.\\
\indent Recall that \[
\frac{\int_{(0,s^*)}x^{\frac{N}{N-1}}dF(x)}{(s^*)^{\frac{N}{N-1}}}=(N-1)(1-F(s^*)).\]
Subtracting $(N-1)s^*f(s^*)$ from both sides, I obtain that 
\[
\frac{\int_{(0,s^*)}x^{\frac{N}{N-1}}dF(x)}{(s^*)^{\frac{N}{N-1}}}-(N-1)s^*f(s^*)=(N-1)[1-F(s^*)-s^*f(s^*)].\]
The L.H.S. of the above equation is weakly negative, shown in  \eqref{b22}. Together with  that $1-F(x)-xf(x)$ is non-increasing, I have that  $1-F(x)-xf(x)\le 0$ for any $x\ge s^*$. Hence, \[\phi_i^{***}(v)=v_i-\frac{1-F(v_i)}{f(v_i)}\ge 0 \quad \text{for $v\in V^+, s^*<v_i=v(1)<\bar{v}$.}\]. \\
\indent Then, any standard dominant-strategy mechanism,  in which 1) the ex-post participation constraints are binding for zero-valuation bidders and 2) the highest bidder with a valuation higher than $s^*$ is allocated with probability one,  maximizes the expected revenue across standard dominant-strategy mechanisms under $\pi^{***}$. And the second-price auction with the $s^*-$scaled $Beta(\frac{1}{N-1},1)$ distributed random reserve is such a mechanism. Finally, it is straightforward to show that the maximized expected revenue is $Ns^*(1-F(s^*))$ by calculating the expected total virtual surplus.  Indeed, under this mechanism and the joint distribution $\pi^{***}$ , the expected total virtual surplus is \[
\begin{split}
  & N \{\int_{(0,\bar{v})}\int_{(0,s^*)}\frac{f(v_1)}{(N-1)s^*(1-F(s^*))}[v_2f(v_2)-\frac{v_2^{-\frac{N}{N-1}}}{N-1}\int_0^{v_2}x^{\frac{N}{N-1}}f(x)dx)]\cdot (v_1-\frac{1-F(v_1)}{f(v_1)})dv_2dv_1 +\\ & \int_{(0,s^*)}\frac{Pr(\bar{v})}{(N-1)s^*(1-F(s^*))}[v_2f(v_2)-\frac{v_2^{-\frac{N}{N-1}}}{N-1}\int_0^{v_2}x^{\frac{N}{N-1}}f(x)dx)] \cdot \bar{v} dv_2\}\\ &=
  \frac{N}{(N-1)s^*(1-F(s^*))}\cdot [\int_{(0,\bar{v})}(v_1f(v_1)-1+F(v_1))dv_1+Pr(\bar{v})\bar{v}]\cdot \frac{\int_{(0,s^*)}s^{\frac{N}{N-1}}f(s)ds}{(s^*)^{\frac{1}{N-1}}}\\ &= \frac{N}{(N-1)s^*(1-F(s^*))}\cdot [\int_{(0,\bar{v})}(v_1f(v_1)-1+F(v_1))dv_1+Pr(\bar{v})\bar{v}]\cdot (N-1)s^*(1-F(s^*))\\ &= N\cdot [\int_{(0,\bar{v})}(v_1f(v_1)-1+F(v_1))dv_1+Pr(\bar{v})\bar{v}]\\ &= N s^*(1-F(s^*)),
\end{split}
\] 
where the first equality follows from \eqref{eq25}, the second equality follows from \eqref{sd} and the last equality uses integration by parts.
\end{proof}
Theorem \ref{t5} follows immediately from Proposition \ref{p9} and Proposition \ref{p10}.
\section{Proofs for Section \ref{s6}}\label{ac}
\subsection{Proof of Proposition \ref{p6}}
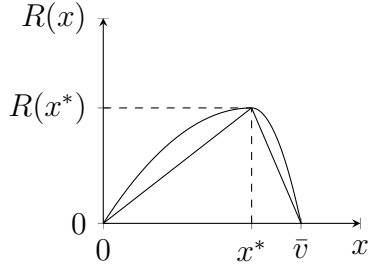
\begin{figure}
\centering
\begin{tikzpicture}
\begin{axis}[
    axis lines = left,
    xmin=0,
        xmax=1.3,
        ymin=0,
        ymax=1,
        xtick={0,0.75,1,1.3},
        ytick={0,0.75^2,1},
        xticklabels = {$0$, $x^*$, $\bar{v}$, $x$},
        yticklabels = {$0$, $R(x^*)$,  $R(x)$},
        legend style={at={(1.1,1)}}
]

\addplot[domain=0:0.75,color=black, name path=X] {-1*(x-0.75)^2+0.75^2};
\addplot[domain=0.75:1,color=black, name path=Y] {-9*(x-0.75)^2+0.75^2};
\addplot[black] coordinates {(0, 0) (0.75,0.75^2)};
\addplot[black] coordinates {(1, 0) (0.75,0.75^2)};
\addplot[dashed] coordinates {(0, 0.75^2) (0.75,0.75^2)};
\addplot[dashed] coordinates {(0.75, 0) (0.75,0.75^2)};
\end{axis}
\end{tikzpicture}
\caption{The curve is a strictly concave revenue function. The L.H.S. of \eqref{d1} is the area under the curve. The R.H.S. of \eqref{d1} is the area of the triangle. } \label{fig:S1}
\end{figure}

Note that  under a posted-price mechanism,  the maximally positively correlated distribution (the valuations of the bidders are always the same) is a worst-case correlation structure, and the revenue guarantee of any posted-price mechanism  is thus at most $\max_{x\in[0,\bar{v}]} x(1-F(x))$,  which is the monopoly profit when there is only one bidder.  It is straightforward to show that  $\frac{x^{\frac{N}{N-1}}}{\bar{v}^{\frac{1}{N-1}}}\ge \frac{x^2}{\bar{v}}$ for any $x\in[0,\bar{v}]$ and $N\ge 2$. Thus,  it suffices to compare
$\frac{E[X^2]}{\bar{v}}$ with $\max_{x\in[0,\bar{v}]} x\cdot (1-F(x))$ if the revenue function $R(x)= x\cdot (1-F(x))$ is strictly concave.  Using integration by parts, I obtain that \[
E[X^2]=2\int_0^{\bar{v}} x(1-F(x))dx.\]
Let $x^*$ denote the unique solution to $\max_{x\in[0,\bar{v}]} R(x)$.  Then using graph (see Figure \ref{fig:S1}) it is straightforward  that
\[\int_0^{\bar{v}}R(x)dx>\frac{1}{2}\cdot \bar{v} \cdot R(x^*).\tag{D.1}\label{d1}\]
\eqref{d1} is equivalent to that
\[
\frac{E[X^2]}{\bar{v}}=\frac{2\int_0^{\bar{v}} x(1-F(x))dx}{\bar{v}} > \max_{x\in [0,\bar{v}]} x(1-F(x)).\]
\subsection{Proof of  Lemma \ref{l1}}
For each $t$, I construct the dual variables for the second-price auction with  the random reserve whose cumulative distribution function is $G_t(r)=(\frac{r}{t})^{\frac{1}{N-1}}$  as follows: \[
\lambda_i(x)=\frac{x^{\frac{N}{N-1}}}{Nt^{\frac{1}{N-1}}} \quad \text{if}\quad 0\le x \le t,\forall i\in I,\]
\[
\lambda_i(x)=\frac{t}{N} \quad \text{if}\quad t< x \le \bar{v},\forall i\in I.\]
Given the constructed dual variables above, the value of $(D)$ is \[\int_{0}^{t}\frac{x^{\frac{N}{N-1}}}{t^{\frac{1}{N-1}}}dF(x)+t(1-F(t)). \] 
Then it suffices to show that the constructed dual variables are feasible, or \eqref{eq27} holds. I divide the valuation profiles into three cases.\\
\textit{Case 1}: $v(1)\le t$.\\
\eqref{eq27} holds by a similar argument with that in the proof of Proposition \ref{p3}.\\
\textit{Case 2}: $v(1)> t, \#\{k:v_k=v(1)\}=1$.\\
When $v(2)> t$, then $t(v)=v(2)$. The L.H.S. of \eqref{eq27} is maximized when $v_i\ge t$ for all $i$, and the maximized value is  $N\cdot \frac{t}{N}=t < t(v)$. When $v(2)\le t$, then $t(v)=v(1)\cdot 1-\int_t^{v(1)}dx-\int_{v(2)}^t(\frac{x}{t})^{\frac{1}{N-1}}dx=\frac{t}{N}+\frac{(N-1)v(2)^{\frac{N}{N-1}}}{Nt^{\frac{1}{N-1}}}$. The L.H.S. of \eqref{eq27} is maximized when $v_i=v(2)$ for all $i\notin \{k:v_k=v(1)\} $, and the maximized value is $\frac{t}{N}+\frac{(N-1)v(2)^{\frac{N}{N-1}}}{Nt^{\frac{1}{N-1}}}=t(v)$. Therefore \eqref{eq27} holds.\\
\textit{Case 3}: $v(1)> t, \#\{k:v_k=v(1)\}\ge 2$.\\
Now $t(v)=v(1)$. The L.H.S. of \eqref{eq27} is maximized when $v_i\ge t$ for all $i$, and the maximized value is $N\cdot \frac{t}{N}=t < t(v)$. Therefore \eqref{eq27} holds.
\subsection{Proof of Theorem \ref{t3}}
Recall that the revenue guarantee of a posted-price mechanism is at most $\max_{x\in[0,\bar{v}]}x\cdot (1-F(x))$ for a given $F\in \Delta^c[0,\bar{v}]$. Denote a solution to $\max_{x\in[0,\bar{v}]}x\cdot (1-F(x))$  as $x^*$. For a given $F\in \Delta^c[0,\bar{v}]$, consider the second-price auction with the $x^*-$scaled $Beta (\frac{1}{N-1},1)$ distributed random reserve, and  I have that
\[
     \int_{0}^{x^*}\frac{x^{\frac{N}{N-1}}}{(x^*)^{\frac{1}{N-1}}}dF(x)+x^*(1-F(x^*)) > x^*(1-F(x^*)),
\]
where the inequality follows from that  $x^*>0$.
\subsection{Proof of Theorem \ref{t4}}
By \cite{he2022correlation}, given a $F\in \Delta^c[0,\bar{v}]$,  the revenue guarantee of the second-price auction with the optimal deterministic reserve for the $N$-bidder case  is as follows: \[
\frac{N}{N-1}\int_{r^*}^{c(r^*)}xdF(x),\] where $r^*$ satisfies $F(Nr^*)=F(\frac{N-1+F(r^*)}{N})$,  and $c(r^*)=F^{-1}(\frac{N-1+F(r^*)}{N})$.\\
\indent Define $J(x):=\frac{N}{N-1}x-\frac{x^{\frac{N}{N-1}}}{c(r^*)^{\frac{1}{N-1}}}$. Because $J'(x)=\frac{N}{N-1}(1-(\frac{x}{c(r^*)})^{\frac{1}{N-1}})$ and $J''(x)=-\frac{Nx^{\frac{1}{N-1}-1}}{(N-1)^2c(r^*)^{\frac{1}{N-1}}}\le 0$, $J(x)$ is maximized at $x=c(r^*)$ and the maximized value is $\frac{1}{N-1}c(r^*)$ by simple calculation. For a given $F\in \Delta^c[0,\bar{v}]$, consider the second-price auction with the $c(r^*)-$scaled $Beta (\frac{1}{N-1},1)$ distributed random reserve, and  I have that
\[
    \begin{split}
        \int_{0}^{c(r^*)}\frac{x^{\frac{N}{N-1}}}{[c(r^*)]^{\frac{1}{N-1}}}dF(x)+c(r^*)[1-F(c(r^*))] & >
        \int_{r^*}^{c(r^*)}\frac{x^{\frac{N}{N-1}}}{[c(r^*)]^{\frac{1}{N-1}}}dF(x)+c(r^*)[1-F(c(r^*))]\\
        & \ge
        \int_{r^*}^{c(r^*)}[\frac{N}{N-1}x-\frac{1}{N-1}c(r^*)]dF(x)+c(r^*)[1-F(c(r^*))]\\
        & =
       \frac{N}{N-1}\int_{r^*}^{c(r^*)}xdF(x),
    \end{split}
\]
where the first inequality follows from that  $r^*>0$, the second inequality follows from that  $J(x)\le \frac{1}{N-1}c(r^*)$ and the equality follows from that  \[
\begin{split}
    \int_{r^*}^{c(r^*)}\frac{1}{N-1}c(r^*)dF(x)&=\frac{1}{N-1}c(r^*)[F(c(r^*))-F(r^*)]\\
   &= \frac{1}{N-1} [\frac{N-1+F(r^*)}{N}-F(r^*)]F^{-1}(\frac{N-1+F(r^*)}{N})\\
   &= [\frac{1-F(r^*)}{N}]F^{-1}(\frac{N-1+F(r^*)}{N})\\ &=
   c(r^*)[1-F(c(r^*))].
\end{split}\]

\section{ ``Necessity'' of Robust Regularity Conditions}\label{ad}
\begin{definition}
\normalfont I say the allocation rule $q$ is \textit{strictly monotone} if for any  $i$, any $v_{-i}$  and any pair of valuation $v_i$ and $v_i'$ in which $q_i(v_i,v_{-i})>0$ and  $q_i(v_i',v_{-i})>0$, I have that  $q_i(v_i,v_{-i})<q_i(v_i',v_{-i})$ whenever $v_i<v_i'$.
\end{definition} 
\begin{proposition}\label{p7}
For the two-bidder case, if the second-price auction with the  uniformly distributed random reserve is a maxmin mechanism across dominant-strategy  mechanisms, then the two-bidder robust regularity conditions hold almost surely. 
\end{proposition}
\begin{proof}
 The intuition behind this is the observation that  under the second-price auction with the uniformly distributed random reserve, the  allocation rule is  strictly monotone. In addition, the high bidder's allocation is positive but less than 1 when  her valuation is positive but less than $\bar{v}$. Thus in a Nash equilibrium, the high bidder's virtual value has to be 0 for these valuations under the joint distribution, otherwise Myerson's ironing argument implies that allocation rule in equilibrium should exhibit ``flatness'' across some range.  Formally, I will establish Lemma \ref{l7} and Lemma \ref{l8} below.
\begin{lemma}\label{l7}
For the two-bidder case, for  the second-price auction with the uniformly distributed random reserve to be part of a Nash equilibrium across dominant-strategy mechanisms, the equilibrium joint distribution has to be $\pi^*$ almost surely.
\end{lemma}
\begin{proof}
 let $\pi$ be a best response of adversarial nature to the second-price auction with the uniformly distributed random  reserve. Suppose \eqref{1} does not hold for a  set of  $(v_1,v_2)$ where $\bar{v}>v_1\ge v_2$ with some positive measure. If virtual values of bidder 1 for these valuation profiles are all positive, then consider a modified allocation exhibiting the property that the allocation to bidder 1 is one from the valuation profile in which the virtual value of bidder 1 becomes positive for the first time. Formally, let $\overline{v_1(v_2)}=\inf\{v_1: \phi_1(v_1,v_2)>0,v_1\ge v_2\}$. Let $\tilde{q}_1(v_1,v_2)=1$ for $v_1>\overline{v_1(v_2)}$ and $\tilde{q}(v):=q^*(v)$ otherwise.  Such modification is feasible since bidder 2 gets zero allocation for any one of  these valuation profiles in the second-price auction with the uniformly distributed random reserve. Thus I have a profitable and feasible deviation.  If virtual values of bidder 1 for these valuation profiles are all negative, by a similar argument, I rule out the possibility that the second-price auction with the uniformly distributed random reserve is a best response of the auctioneer to $\pi$. Now If virtual values of bidder 1 for these valuation profiles are not all positive and not all negative, I have to discuss two cases. The first case is that the virtual value is still (weakly) monotone. Then by a similar argument, the second-price auction with the  uniformly distributed random reserve can not a best response to $\pi$. The second case is that the virtual value is not monotone, then a best response has to exhibit flatness across a range of valuation profiles, which can be done by Myerson's ironing procedure. Recall that the allocation rule of the second-price auction with the uniformly distributed random reserve is strictly monotone. Thus, it cannot be a best response. To illustrate this, suppose $\phi_1(\cdot,v_2)$ is decreasing in $v_1$ for  $v_1\in (a(v_2),b(v_2))$ and $\phi_1(a(v_2),v_2)>\phi_1(\hat{v}_1(v_2),v_2)=0>\phi_1(b(v_2),v_2)$ for some $\hat{v}_1(v_2)\in (a(v_2),b(v_2))$. Then let $\tilde{q}_1(v_1,v_2)=q^*(\hat{v}_1(v_2),v_2)$ for $v_1\in [a(v_2),b(v_2)]$ and $\tilde{q}(v)=q^*(v)$ otherwise. Since this is a feasible and  profitable deviation, I conclude that the second-price auction with the uniformly distributed random reserve can not be a best response.\\
\indent Together with the proof of Proposition \ref{p3}, the equilibrium joint distribution is $\pi^*$ almost surely.
 \end{proof}
 \begin{lemma}\label{l8}
For the two-bidder case,  for the second-price auction with the uniformly distributed random reserve to be part of a Nash equilibrium across dominant-strategy mechanisms, $\pi^*$ exhibits \eqref{2} almost surely.
 \end{lemma}
 \begin{proof}
  Suppose not. Then, there exists a set of  $(v_1, v_2)$ where $0<v_2<v_1<\bar{v}$ but $\phi_2(v_1, v_2)>0$ with some positive measure. Then  by increasing the allocation to bidder 2 by a small amount $\epsilon$ when the valuation profile lies in this set, I have a feasible and profitable deviation. Thus,  the second-price auction with the uniformly distributed random reserve is not a best response. 
 \end{proof}
 Proposition \ref{p7} follows immediately from  Lemma \ref{l7}, Lemma \ref{l8} and the proof of Proposition \ref{p3}.
 \end{proof}
\begin{proposition}\label{p8}
For the $N-$bidder ($N\ge 3$)  case,   If the second-price auction with the $\bar{v}-$scaled $Beta (\frac{1}{N-1},1)$ distributed random reserve is a maxmin mechanism across standard dominant-strategy  mechanisms, then the $N-$bidder robust regularity conditions (I) hold almost surely. 
\end{proposition}
\begin{proof}
First, I establish Lemma \ref{l15} below.
\begin{lemma}\label{l15}
For the $N-$bidder ($N\ge 3$)  case, for the second-price auction with the $Beta(\frac{1}{N-1},1)$ distributed random reserve to be part of a Nash equilibrium across standard dominant-strategy mechanisms, the equilibrium joint distribution has to be $\pi^{**}$ almost surely.
\end{lemma}
\begin{proof}
As shown in the proof of Proposition  \ref{p4}, \eqref{eq27} holds with equality if and only if $v\in V^+$. This implies the equilibrium joint distribution has the support $V^+$. Then this lemma follows from  a similar argument to the proof of Lemma \ref{l7}. 
\end{proof}
Proposition \ref{p8} follows immediately from  Lemma \ref{l15} and the proof of Proposition \ref{p5}.
\end{proof}

%
%

\end{document}